\documentclass[amsfonts,amsmath,prd,preprint, nofootinbib]{revtex4}

\usepackage{epsfig,bbm,cancel,ulem}
\usepackage[breaklinks=true]{hyperref}
\usepackage{xcolor}
\usepackage{wasysym}
\usepackage[mathscr]{eucal}
\usepackage{multirow}
\usepackage{amsmath}
\usepackage{enumitem}
\usepackage[sort&compress]{natbib}

\begin{document}

\title{Momentum Diffusion, Decoherence and Drag Force on a Magnetic Nanoparticle}

\author{Agya Sewara Alam}
\email{a.s.alam@rug.nl}
\affiliation{Van Swinderen Institute for Particle Physics and Gravity, University of Groningen, 9747AG Groningen, the Netherlands}
\affiliation{Departemen Fisika, FMIPA, Universitas Indonesia, Depok, 16424, Indonesia.}

\author{Anupam Mazumdar}
\affiliation{Van Swinderen Institute for Particle Physics and Gravity, University of Groningen, 9747AG Groningen, the Netherlands.}

\begin{abstract}

In this paper, we will provide a complete derivation of the decoherence rate for a magnetic nanoparticle in quantum superposition in the presence of the fluctuating electromagnetic field in a thermal background by using the fluctuation-dissipation theorem in the long-wavelength limit. The long-wavelength limit assumes that the superposition size is much smaller than the wavelength of the electromagentic filed fluctuations. We will extend this computation to two diamagnetic nanoparticles kept in quantum superposition adjacent to each other.
We will also show how the drag force on a single nanoparticle arises from external electromagnetic-field fluctuations, and compare our results with those for the nanoparticle's dielectric properties. 
\end{abstract}

\maketitle
\thispagestyle{empty}
\setcounter{page}{1}

\section{Introduction}
\label{sec:intro}

Diamagnetic levitation of a nanoparticle is one of the promising experimental platform to test many fundamental physics, with a multitude of applications in classical ~\cite{Simon:2001,simon_diamagnetic_2000}, and quantum experiments, especially, in the context of quantum sensors~\cite{Hofer:2022chf}, quantum metrology~\cite{Nakashima:2020jnr,chen2022diamagnetic}, and schemes for matter-wave interferometery~\cite{Marshman:2018upe,PhysRevLett.125.023602,marshman_constructing_2022}, for a review see~\cite{schilling2021physicsdiamagneticlevitation}. The latter application provides the novelty as one can test the foundations of quantum mechanics, such as how heavy a macroscopic quantum superposition we can create in our lab, and second, we can potentially test an ambitious program to test the quantum nature of gravity in a lab via quantum gravity induced entanglement of matter (QGEM) protocol~\cite{Bose:2017nin,ICTS}, see also~\cite{Marletto:2017kzi}. The authors of \cite{Bose:2017nin} proposed  
a concrete experimental setup where a nanodiamond will be levitated diamagnetically~\cite{Schut:2021svd,Elahi:2024dbb}, and the embedded spin of the nitrogen vacancy (NV) center can be used to create a Stern-Gerlach type experiment to create a spatial superposition, see~\cite{marshman_constructing_2022}, similar to the atom interferometer case~\cite{Folman:2013,PhysRevLett.123.083601,doi:10.1126/sciadv.abg2879}.

These quantum experiments require stringent constraints 
arising from decoherence~\cite{romero-isart_quantum_2011,RomeroIsart2011LargeQS,schlosshauer_quantum_2019,Schut:2024lgp} from the ambient vacuum~\cite{van_de_kamp_quantum_2020,Schut:2023eux,Schut:2021svd,Rijavec:2020qxd}, external dipoles~\cite{Fragolino:2023agd}, external electromagnetic fluctuations due to current and external charge/neutral particles~\cite{Schut:2023tce,Moorthy:2025fnu, Moorthy:2025bpz}, external jitters and relative acceleration noise~\cite{Toros:2020dbf,Wu:2024bzd}.

In general, the coherence of the levitating nanoparticle will also be sensitive to the electromagnetic vacuum fluctuations present inside the experimental box. These fluctuations are not zero-temperature but finite-temperature fluctuations, see~\cite{Sinha:2022snc}. Analysing these effects, including decoherence and the drag force due to external electromagnetic fluctuations acting on the nanoparticle, is of paramount importance for any quantum spatial superposition experiment. Recently, this issue was addressed in \cite{Zhang:2025fxs}, but the results were limited to magnetic-field fluctuations and did not include time-dependent electric-field fluctuations associated with the electromagnetic fluctuations. As it turns out, time-dependent electric-field fluctuations play a significant role in decoherence, as we will show in this paper. Also, the paper \cite{Zhang:2025fxs} did not compute the thermal drag on a diamagnetic nanoparticle. Here, we will compute this effect first in a relativistic setting and then take the non-relativistic limit following the arguments first proposed in \cite{Sinha:2022snc}. However, the authors computed the drag force due to vacuum fluctuations for a nanoparticle's dielectric properties.

Analysis and results of this paper are motivated by the earlier results of this very interesting research article ~\cite{Sinha:2022snc}~\footnote{Previously, the study of electromagnetic fluctuation on a metallic material has been analysed \cite{rytov_theory_1959,zheng_review_2014,brevik_fluctuational_2022}. However, Ref.~\cite{Sinha:2022snc} had examined it more comprehensively by including momentum diffusion, decoherence rate, and drag force in the context of dissipation fluctuation analysis.}.
The authors computed the decoherence effect on a dielectric nanoparticle, derived from the momentum diffusion constant in the presence of blackbody radiation, and they demonstrated that it is closely related to the long-wavelength collisional decoherence rate for such a nanoparticle in a thermal environment~\cite{schlosshauer,Hornberger:2003,Schut:2024lgp}. The authors~\cite{Sinha:2022snc} also showed that the diffusion constant appears in the steady-state photon emission rate of two dipoles induced by blackbody radiation, and also computed the Einstein--Hopf drag force on a small polarizable nanoparticle moving in a blackbody field. Following the results of this paper~\cite{Sinha:2022snc}, we now embark on computing the same for a diamagnetic nanoparticle, and we will comment on the nitrogen-vacancy-centred nanodiamond's properties. We wish to employ the fluctuation dissipation analysis~\cite{Fokker:1914,Berg-Sorenson_1992,Balykin_1986,Dalibard_1985,Agarwal:1993} to study how the decoherence rate and drag force are compared to those of the dielectric scenario studied already in \cite{Sinha:2022snc}. 

In this paper, we will compute the decoherence rate in the long-wavelength limit for magnetic dipole moments in a magnetic field and a time-varying electric field. We will show that the dominant contribution to the momentum diffusion constant will arise from the time-varying electric field on the magnetic dipoles. 
There will be three contributions to momentum variance: magnetic, electric, and the coupled term between magnetic and electric fields.
We will compute the decoherence rate in the presence of another dipole, and finally, we will compute the drag force on the diamagnetic nanoparticle following the computations of \cite{Sinha:2022snc}.

We organise the paper as follows. In Section \ref{sec:mdc}, we derive the momentum diffusion constant of the particles with a magnetic field and a time-dependent electric field.  We also use the independence of the magnetic field from its spatial derivative, as in Einstein and Hopf \cite{Einstein:1910a} \cite{Einstein:1910b}. 
In Section \ref{sec:decoherence}, we derive the decoherence rate from the photon emission rate between two magnetic dipoles within a magnetic field. We show the decoherence rate of a diamagnetic nanosphere in long-wavelength approximation and compare it with the dielectric one. In Section \ref{sec:dragforce}, we will derive the drag force of a moving magnetic dipole within a magnetic field in relativistic form. Then, we take the non-relativistic limit and compare the diamagnetic and dielectric nanospheres.

\section{Momentum Diffusion Constant}
\label{sec:mdc}

Let us consider a nanosphere of radius $a$, with a point-like magnetic dipole moment~\footnote{We can justify the point-like dipole moment if the external magnetic field is nearly constant over the size of the nanosphere, see~\cite{Elahi:2024dbb}.}. We will generalise our results to finite volumes by employing the Clausius-Mossotti relation. Let us consider the force acting on a diamagnetic nanosphere due to external magnetic and electric fields, also writen in the component form,\cite{Zangwill_2012}:
\begin{equation}\label{eq:fx}
   \textbf{F} = \nabla(\textbf{m} \cdot \textbf{B}) + \frac{1}{c^2} \textbf{m} \times \partial_t \textbf{E}\,,~~~~  F_i = m_j \partial_i B_j + \frac{1}{c^2} \epsilon_{ijk} m_j \partial_t E_k.
\end{equation}
where $\textbf{m}$ is the magnetic dipole moment. 
Where  $k=(x,y,z)$, respectively, and $\epsilon_{ijk}$ is Levi-Civita~\footnote{ In the previous study~\cite{Zhang:2025fxs}, we neglected the second contribution in Eq.~(\ref{eq:fx}). However, it turns out that the second contribution dominates over the first, and here we compare the results of the two papers. } \footnote{A steady electric field will not exert force on a magnetic dipole. However, a time-dependent electric field induces a magnetic field through Maxwell-Ampere law, $\nabla \times \textbf{B}=\frac{1}{c^2} \frac{\partial \textbf{E}}{\partial t}$, so it would contribute to a force acting on a magnetic dipole.}. The magnetic dipole moment can be written in
   $ m_j = \alpha B_j$,
where $\alpha$ depends on the magnetic polarizability. For an NV-centered nanodiamond, $\alpha $ is a complex entity and depends on the frequency and the electronic spin resonance of the NV spin,~see~\cite{Eisenach:2021}.
Now, let us assume that the nanodiamond is present in the external electric the magnetic fields, which can be recast in terms  of the creation and annihilation operators 
\begin{eqnarray}
    B_j &=& i \sum_{\textbf{k} \lambda}  \left( \frac{\hbar}{2 \varepsilon_0 \omega V} \right)^{1/2} \left[ a_{\textbf{k} \lambda} e^{i (\textbf{k} \cdot \textbf{x} - \omega t)} - a_{\textbf{k} \lambda}^{\dagger} e^{-i (\textbf{k} \cdot \textbf{x} - \omega t)} \right] (\epsilon_{jik} \textbf{k}_i \textbf{e}_{\textbf{k} \lambda,k}), \label{eq:B} \\
    E_k &=& i \sum_{\textbf{k} \lambda}  \left( \frac{\hbar \omega}{2 \varepsilon_0 V} \right)^{1/2} \left[ a_{\textbf{k} \lambda} e^{i (\textbf{k} \cdot \textbf{x} - \omega t)} - a_{\textbf{k} \lambda}^{\dagger} e^{-i (\textbf{k} \cdot \textbf{x} - \omega t)} \right] \textbf{e}_{\textbf{k} \lambda, k}. \label{eq:E}
\end{eqnarray}
Where $\textbf{e}_{\textbf{k} \lambda, k}$ is the polarization vector unit in $k-$direction. The $a_{\textbf{k} \lambda},~a^{\dagger}_{\textbf{k} \lambda}$ are the photon annihilation and the creation operators.  Their commutation relations are given by:
\begin{eqnarray}
    [a_{\textbf{k} \lambda},a^{\dagger}_{\textbf{k}' \lambda}] &=& \delta_{\textbf{kk}'} \delta_{\lambda \lambda'},~~~~~~~~~
    \left[ a_{\textbf{k} \lambda},a_{\textbf{k}' \lambda} \right] = 0.
\end{eqnarray}
The derivatives of $B_j,~E_k$ with respect to $x$ and time $t$, respectively, are given by:
\begin{eqnarray}
\label{eq:partialder}
    \partial_i B_j &=& - \sum_{\textbf{k} \lambda}  \left( \frac{\hbar}{2 \varepsilon_0 \omega V} \right)^{1/2} k_i \left[ a_{\textbf{k} \lambda} e^{i (\textbf{k} \cdot \textbf{x} - \omega t)} + a_{\textbf{k} \lambda}^{\dagger} e^{-i (\textbf{k} \cdot \textbf{x} - \omega t)} \right] (\epsilon_{jik} \textbf{k}_i \textbf{e}_{\textbf{k} \lambda,k}), \\
    \partial_t E_k &=& \sum_{\textbf{k} \lambda}  \left( \frac{\hbar \omega}{2 \varepsilon_0 V} \right)^{1/2} \omega \left[ a_{\textbf{k} \lambda} e^{i (\textbf{k} \cdot \textbf{x} - \omega t)} + a_{\textbf{k} \lambda}^{\dagger} e^{-i (\textbf{k} \cdot \textbf{x} - \omega t)} \right] \textbf{e}_{\textbf{k} \lambda,k}.
\end{eqnarray}
Hence, the magnetic dipole moment can be written as
\begin{equation} \label{eq:mj}
    m_j = i \sum_{\textbf{k} \lambda}  \left( \frac{\hbar}{2 \varepsilon_0 \omega V} \right)^{1/2} \left[ \alpha(\omega)a_{\textbf{k} \lambda} e^{i (\textbf{k} \cdot \textbf{x} - \omega t)} - \alpha^*(\omega) a_{\textbf{k} \lambda}^{\dagger} e^{-i (\textbf{k} \cdot \textbf{x} - \omega t)} \right] (\epsilon_{jik} \textbf{k}_i \textbf{e}_{\textbf{k} \lambda,k}),
\end{equation}
where $\alpha(\omega)$ is the magnetic susceptibility of the magnetic dipole moment, which has a frequency dependence. Therefore, Eq.~(\ref{eq:fx}) can be written as
\begin{eqnarray} \label{eq:pgeneral}
\Delta p_i &= &\int_{0}^{\Delta t} dt F_i\nonumber = \int_0^{\Delta t} dt m_j \partial_i B_j + \frac{1}{c^2} \epsilon_{ijk} \int_0^{\Delta t} dt m_j \partial_t E_k, \nonumber \\
    \Delta p_i &=& \Delta p_M + \Delta p_E,
\end{eqnarray}
where the $\Delta p_M$ and $\Delta p_E$ are the momenta affected by spatial variation of the magnetic and the time-dependent part of the electric fields, respectively. In this calculation, we include the fluctuations of both the fields, where $\Delta p_E$ was ignored in Ref.~\cite{Zhang:2025fxs}, while the magnetic term can be obtained correctly as~\cite{Zhang:2025fxs}:
\begin{eqnarray}
\label{eq:mag}
    \Delta p_M &=& -2i \sum_{\textbf{k}_1 \lambda_1} \sum_{\textbf{k}_2 \lambda_2}  \left( \frac{\hbar}{2 \varepsilon_0 \omega_1 V} \right)^{1/2} \left( \frac{\hbar}{2 \varepsilon_0 \omega_2 V} \right)^{1/2} k_{1,i} \frac{\sin{[\frac{1}{2} (\omega_1 - \omega_2)\Delta t]}}{\omega_1 - \omega_2} \times \nonumber \\
    & &  \left[ \alpha(\omega_2) a_{\textbf{k}_2 \lambda_2} a_{\textbf{k}_1 \lambda_1}^{\dagger} \xi e^{i (\omega_1 - \omega_2) \Delta t /2} - \alpha^*(\omega_2)a_{\textbf{k}_2 \lambda_2}^{\dagger} a_{\textbf{k}_1 \lambda_1} \xi^* e^{-i (\omega_1 - \omega_2) \Delta t /2} \right] \times \nonumber \\
    & & (\epsilon_{jik} \textbf{k}_{1,i} \textbf{e}_{\textbf{k}_1 \lambda_1,k}) (\epsilon_{jik} \textbf{k}_{2,i} \textbf{e}_{\textbf{k}_2 \lambda_2 ,k}),
\end{eqnarray}
where indices $i,j,k =1,2,3$ and $\xi=e^{i(\textbf{k}_2 \cdot \textbf{x}_2 -\textbf{k}_1 \cdot \textbf{x}_1)}$. The electric term, where the details are in appendix \ref{sec:pE}, can be derived as 
\begin{eqnarray} \label{eq:deltapE}
    \Delta p_E &=& \frac{2i}{c^2} \sum_{\textbf{k}_1 \lambda_1} \sum_{\textbf{k}_2 \lambda_2}  \left( \frac{\hbar \omega_1}{2 \varepsilon_0 V} \right)^{1/2} \left( \frac{\hbar}{2 \varepsilon_0 \omega_2 V} \right)^{1/2} \omega_{1} \frac{\sin{[\frac{1}{2} (\omega_1 - \omega_2)\Delta  t]}}{\omega_1 - \omega_2} \times \nonumber \\
    & &  \left[ \alpha(\omega_2) a_{\textbf{k}_2 \lambda_2} a_{\textbf{k}_1 \lambda_1}^{\dagger} \xi e^{i (\omega_1 - \omega_2) \Delta t /2} - \alpha^*(\omega_2) a_{\textbf{k}_2 \lambda_2}^{\dagger} a_{\textbf{k}_1 \lambda_1} \xi^* e^{-i (\omega_1 - \omega_2) \Delta t /2} \right] \times \nonumber \\
    & & ( \epsilon_{ijk} \textbf{e}_{\textbf{k}_1 \lambda_1 ,k}) (\epsilon_{jik} \textbf{k}_{2,i}\textbf{e}_{\textbf{k}_2 \lambda_2 ,k}).
\end{eqnarray}
The expectation value of momentum can be written as:
\begin{eqnarray}\label{totalp}
    \Delta p^2 &=& \Delta p^2_M + \Delta p_E^2 + \Delta p_M^{\dagger} \Delta p_E + \Delta p_E^{\dagger} \Delta p_M =
  \langle \Delta p^2_M \rangle + \langle \Delta p_E^2 \rangle + \langle \Delta p_c^2 \rangle,
\end{eqnarray}
where we define $$\langle \Delta p_c^2 \rangle = \langle \Delta p_M^{\dagger} \Delta p_E \rangle + \langle \Delta p_E^{\dagger} \Delta p_M \rangle $$. The magnetic and electric momentum squared terms can be written as:
\begin{eqnarray}
    \Delta p_M^2 &=& 4 \sum_{\textbf{k}_1 \lambda_1} \sum_{\textbf{k}_2 \lambda_2}  \left( \frac{\hbar}{2 \varepsilon_0 \omega_1 V} \right) \left( \frac{\hbar}{2 \varepsilon_0 \omega_2 V} \right) k_{1,i}^2 \frac{\sin^2{[\frac{1}{2} (\omega_1 - \omega_2)\Delta t]}}{(\omega_1 - \omega_2)^2} \times \nonumber \\
    & & |\alpha(\omega_2)|^2 \left[ a_{\textbf{k}_2 \lambda_2}^{\dagger} a_{\textbf{k}_1 \lambda_1} a_{\textbf{k}_2 \lambda_2} a_{\textbf{k}_1 \lambda_1}^{\dagger} + a_{\textbf{k}_2 \lambda_2} a_{\textbf{k}_1 \lambda_1}^{\dagger} a_{\textbf{k}_2 \lambda_2}^{\dagger} a_{\textbf{k}_1 \lambda_1} \right]  (\epsilon_{jik} \textbf{k}_{1,i} \textbf{e}_{\textbf{k}_1 \lambda_1,k})^2 \times \nonumber \\
    & & (\epsilon_{jik} \textbf{k}_{2,i} \textbf{e}_{\textbf{k}_2 \lambda_2,k})^2, \\
    \Delta p_E^2 &=& \frac{4}{c^4} \sum_{\textbf{k}_1 \lambda_1} \sum_{\textbf{k}_2 \lambda_2}  \left( \frac{\hbar \omega_1}{2 \varepsilon_0 V} \right) \left( \frac{\hbar}{2 \varepsilon_0 \omega_2 V} \right) \omega_{1}^2 \frac{\sin^2{[\frac{1}{2} (\omega_1 - \omega_2)\Delta t]}}{(\omega_1 - \omega_2)^2} \times \nonumber \\
    & &  |\alpha(\omega_2)|^2 \left[ a_{\textbf{k}_2 \lambda_2}^{\dagger} a_{\textbf{k}_1 \lambda_1} a_{\textbf{k}_2 \lambda_2} a_{\textbf{k}_1 \lambda_1}^{\dagger} + a_{\textbf{k}_2 \lambda_2} a_{\textbf{k}_1 \lambda_1}^{\dagger} a_{\textbf{k}_2 \lambda_2}^{\dagger} a_{\textbf{k}_1 \lambda_1} \right] ( \epsilon_{ijk} \textbf{e}_{\textbf{k}_1 \lambda_1 ,k})^2 \nonumber \\
    && (\epsilon_{jik} \textbf{k}_{2,i} \textbf{e}_{\textbf{k}_2 \lambda_2 ,k})^2.
\end{eqnarray}
We can now find the thermal expectation value of the fields by multiplying by a $1/3$ factor to account for all field directions and by including finite-temperature effects. The latter effect can be taken into account by taking the finite temperature occupation number of the photons, $n(\omega)$, we get:
\begin{eqnarray}
    \langle a_{\textbf{k} \lambda} a_{\textbf{k} \lambda}^{\dagger} \rangle = n(\omega) + 1,~~~~
    \langle a_{\textbf{k} \lambda}^{\dagger} a_{\textbf{k} \lambda} \rangle = n(\omega),~~~~~
    \omega &=& |\textbf{k}| c,~~~~~
     n(\omega) = \frac{1}{e^{\hbar \omega / k_B T}-1}.
    \label{occupation number}
\end{eqnarray}
where $n(\omega)$ follows the Bose-Einstein statistics, and $k_B$ is the Boltzmann constant.
Hence, the finite temperature expectation values can be written with the help of Eq.~\ref{occupation number}
\begin{eqnarray}
    \langle \Delta p_M^2 \rangle &=& \frac{1}{3} \frac{ \hbar^2}{ \varepsilon_0^2 V^2} \left[ \frac{V}{(2 \pi)^3} \right]^2 \int d^3 k_1 \int d^3 k_2 \frac{k_{1,i}^2}{\omega_1 \omega_2} \frac{\sin^2{[\frac{1}{2} (\omega_1 - \omega_2)\Delta t]}}{(\omega_1 - \omega_2)^2} k_1^2 k_2^2 \times \nonumber \\
    & & |\alpha(\omega_2)|^2 \left[ \langle a_{\textbf{k}_2 \lambda_2}^{\dagger} a_{\textbf{k}_2 \lambda_2} \rangle \langle a_{\textbf{k}_1 \lambda_1} a_{\textbf{k}_1 \lambda_1}^{\dagger} \rangle + \langle a_{\textbf{k}_2 \lambda_2} a_{\textbf{k}_2 \lambda_2}^{\dagger} \rangle \langle a_{\textbf{k}_1 \lambda_1}^{\dagger} a_{\textbf{k}_1 \lambda_1} \rangle \right] \sum_{\lambda_1} (\textbf{e}_{\textbf{k}_1 \lambda_1}^2) \sum_{\lambda_2} ( \textbf{e}_{\textbf{k}_2 \lambda_2}^2), \nonumber \\
    &=& \frac{2 \mu_0^2 \hbar^2}{9 \pi^4 c^{8}} \int_0^{\infty} d\omega \omega^8 |\alpha(\omega)|^2 [n^2(\omega) + n(\omega)] \frac{\pi \Delta t}{2}. \label{eq:pE} \\
    \langle \Delta p_E^2 \rangle &=& \frac{1}{3} \frac{ \hbar^2}{\varepsilon_0^2 c^4 V^2} \left[ \frac{V}{(2 \pi)^3} \right]^2 \int d^3 k_1 \int d^3 k_2 \frac{\omega_1^3}{\omega_2} \frac{\sin^2{[\frac{1}{2} (\omega_1 - \omega_2)\Delta t]}}{(\omega_1 - \omega_2)^2} k_2^2 \times \nonumber \\
    & &  |\alpha(\omega_2)|^2 \left[ \langle a_{\textbf{k}_2 \lambda_2}^{\dagger} a_{\textbf{k}_2 \lambda_2} \rangle \langle a_{\textbf{k}_1 \lambda_1} a_{\textbf{k}_1 \lambda_1}^{\dagger} \rangle + \langle a_{\textbf{k}_2 \lambda_2} a_{\textbf{k}_2 \lambda_2}^{\dagger} \rangle \langle a_{\textbf{k}_1 \lambda_1}^{\dagger} a_{\textbf{k}_1 \lambda_1} \rangle \right] \sum_{\lambda_1} ( \textbf{e}_{\textbf{k}_1 \lambda_1}^2) \sum_{\lambda_2} (\textbf{e}_{\textbf{k}_2 \lambda_2}^2), \nonumber \\
    &=& \frac{2 \mu_0^2 \hbar^2}{3 \pi^4 c^{8}} \int_0^{\infty} d\omega \omega^8 |\alpha(\omega)|^2 [n^2(\omega) + n(\omega)] \frac{\pi \Delta t}{2}\,, \label{eq:pM}
\end{eqnarray}
where the $1/3$ factor appeared in the usual sense of accounting for all three spatial directions~\footnote{
Where we evaluate the integration by using:
 $   \int_0^{\infty} d \omega_2 \frac{\sin^2{[\frac{1}{2} (\omega_1 - \omega_2)\Delta t]}}{(\omega_1 - \omega_2)^2} = \frac{\pi \Delta t}{2}$, the sum of the polarization vector used in this calculation is:
$    \left(\sum_{\lambda_1} \textbf{e}_{\textbf{k}_1 \lambda_1}^2\right)\left(\sum_{\lambda_2} \textbf{e}_{\textbf{k}_2 \lambda_2}^2\right) = 4$, and taking the continuous limit, i.e. $\sum_{\textbf{k}}\rightarrow \frac{V}{(2\pi)^3}\int_0^{\infty}d^3k$.}.
Furthermore, the expectation value of the momentum of the coupled term can be obtained using a similar fashion,
\begin{eqnarray}
    \Delta p_M^{\dagger} \Delta p_E &=& -4\frac{1}{c^2} \sum_{\textbf{k}_1 \lambda_1} \sum_{\textbf{k}_2 \lambda_2} \left( \frac{\hbar}{2 \varepsilon_0 \omega_1 V} \right)^{1/2} \left( \frac{\hbar}{2 \varepsilon_0 \omega_2 V} \right) \left( \frac{\hbar \omega_1}{2 \varepsilon_0 V} \right)^{1/2} k_{1,i} \omega_1 \frac{\sin^2{[\frac{1}{2}(\omega_1 - \omega_2) \Delta t]}}{(\omega_1 - \omega_2)^2} \times \nonumber \\
    & &  |\alpha(\omega_2)|^2 \left[ a_{\textbf{k}_2 \lambda_2}^{\dagger} a_{\textbf{k}_1 \lambda_1} a_{\textbf{k}_2 \lambda_2} a_{\textbf{k}_1 \lambda_1}^{\dagger} + a_{\textbf{k}_2 \lambda_2} a_{\textbf{k}_1 \lambda_1}^{\dagger} a_{\textbf{k}_2 \lambda_2}^{\dagger} a_{\textbf{k}_1 \lambda_1} \right] (\epsilon_{jik} \textbf{k}_{1,i} \textbf{e}_{\textbf{k}_1 \lambda_1 ,k})  (\epsilon_{ijk} \textbf{e}_{\textbf{k}_1 \lambda_1 ,k}) \times \nonumber \\
    & & (\epsilon_{jik} \textbf{k}_{2,i} \textbf{e}_{\textbf{k}_2 \lambda_2 ,k})^2,
\end{eqnarray}
and $\Delta p_E^{\dagger} \Delta p_M = \Delta p_M^{\dagger} \Delta p_E$. Therefore, using a similar calculation, the finite temperature expectation value can be written as
\begin{eqnarray}
\label{eq:pc}
    \langle \Delta p_c^2 \rangle &=& -\frac{2 \hbar^2}{3 (2\pi)^6 \varepsilon_0^2 c^2} \int d^3k_1 \int d^3k_2 \frac{k_{1,i}^2 k_2^2 \omega_1}{\omega_2} \frac{\sin^2{[\frac{1}{2}(\omega_1 - \omega_2) \Delta t]}}{(\omega_1 - \omega_2)^2} \times \nonumber \\
    & & |\alpha(\omega_2)|^2 \left[ \langle a_{\textbf{k}_2 \lambda_2}^{\dagger} a_{\textbf{k}_2 \lambda_2} \rangle \langle a_{\textbf{k}_1 \lambda_1} a_{\textbf{k}_1 \lambda_1}^{\dagger} \rangle + \langle a_{\textbf{k}_2 \lambda_2} a_{\textbf{k}_2 \lambda_2}^{\dagger} \rangle \langle a_{\textbf{k}_1 \lambda_1}^{\dagger} a_{\textbf{k}_1 \lambda_1} \rangle \right] \sum_{\lambda_1} (\textbf{e}_{\textbf{k}_1 \lambda_1}^2) \sum_{\lambda_2} ( \textbf{e}_{\textbf{k}_2 \lambda_2}^2) \nonumber \\
    &=& -\frac{4 \mu_0^2 \hbar^2}{9 \pi^4 c^{8}} \int_0^{\infty} d \omega \omega^8 |\alpha(\omega)|^2 [n^2(\omega) + n(\omega)] \frac{\pi \Delta t}{2}.
\end{eqnarray}
Therefore, the time-derivative of the expectation value of the total momentum, see~\ref{totalp}, can be written as, see~\cite{campbell:01,Sinha:2022snc}
\begin{eqnarray} \label{eq:psq}
     \frac{\langle \Delta p^2 \rangle}{\Delta t}  = \frac{2 \hbar^2 \mu_0^2}{9 \pi^3 c^{8}} \int_0^{\infty} d \omega \omega^8 |\alpha(\omega)|^2 [n^2(\omega) + n(\omega)].  
\end{eqnarray}
The above expression relates to the momentum diffusion constant, which we will discuss below. However, first, we discuss the magnetic polarizability associated with the nanoparticle:
$$\alpha(\omega)=\alpha_{R}(\omega)+i\alpha_{I}(\omega),$$ which has the real and the imaginary part. It was noted in Refs.~\cite{Neuringer1966,stratton2007electromagnetic} that in the Rayleigh scattering limit of a photon with a nanoparticle, the extinction and scattering cross sections are directly related to the imaginary part of the magnetic polarizability $\alpha_I(\omega)\propto|\alpha(\omega)|^2$, when there is no absorption.  In particular, see Appendix \ref{sec:Extinction}, the imaginary part of magnetic polarizability can be written as~\cite{Neuringer1966,stratton2007electromagnetic}:
\begin{equation}\label{eq:noabs}
    \alpha_I (\omega) = \frac{\mu_0}{6\pi} \left( \frac{\omega}{c} \right)^3 |\alpha(\omega)|^2 \beta.
\end{equation}
Where $\beta$ is dimensionless shape parameter; $\beta=1/2$ for a sphere \cite{Neuringer1966}. Therefore, by substituting Eq. (\ref{eq:noabs}) to Eq. (\ref{eq:psq}), we obtain:
\begin{equation}
    \frac{\langle \Delta p^2 \rangle}{\Delta t}  = \frac{4 \hbar^2 \mu_0}{3 \pi^2 c^{5}\beta} \int_0^{\infty} d \omega \omega^5 \alpha_I(\omega) [n^2(\omega) + n(\omega)]. 
\end{equation}
However, this expression includes only the scattering contribution to the imaginary part of the magnetic polarizability and neglects material absorption. However, absorption provides an additional contribution to $\alpha_I(\omega)$. Hence, we should add the extra absorption term into the Eq.~(\ref{eq:noabs}), see~\cite{vanDeHulst1957,bohren1983absorption}, and Appendix~\ref{sec:Extinction}:
\begin{equation}\label{eq:withabs}
    \alpha_I (\omega) = \frac{\mu_0}{6\pi} \left( \frac{\omega}{c} \right)^3 |\alpha(\omega)|^2\beta + \alpha_I^{\mathrm{abs}}(\omega).
\end{equation}
Only the imaginary part of the magnetic polarizability appears explicitly in Eqs. (\ref{eq:noabs})-(\ref{eq:withabs}). This is a direct consequence of the optical theorem, see~\cite{MILONNI_2023}, which relates extinction to the imaginary part of $\alpha(\omega)$. The real part of the $\alpha(\omega)$ contributes only implicitly through the scattering term via $|\alpha(\omega)|^2$.
Therefore, Eq.~(\ref{eq:psq}) can be written as
\begin{eqnarray} \label{eq:genptot}
     \frac{\langle \Delta p^2 \rangle}{\Delta t}  = \frac{4\mu_0 \hbar^2}{3 \pi^2 c^{5}} \int_0^{\infty} d \omega \omega^5 \left[ \mu_0 \frac{\omega^3}{6\pi c^3}|\alpha(\omega)|^2 + \frac{1}{\beta}\alpha_I^{\mathrm{abs}}(\omega) \right] [n^2(\omega) + n(\omega)].  
\end{eqnarray}
This result represents the combined effect of momentum influenced by both magnetic and electric fields, along with their interdependence. We also have a relation from Eqs. (\ref{eq:pE}),(\ref{eq:pM}), and (\ref{eq:pc}) 
\begin{equation}
    |\langle \Delta p^2_E \rangle| > |\langle \Delta p^2_c \rangle| > |\langle \Delta p^2_M \rangle|.
\end{equation}
From Eq.~\ref{eq:genptot}, we see that the time-dependent electric-field-fluctuation term is the dominant one. Therefore, the contribution to the second term of Eq.~(\ref{eq:pgeneral}), which was missed in our earlier paper \cite{Zhang:2025fxs}, is more dominant than the magnetic part. Since the magnetic term has the smallest contribution, the absolute coupled term is larger. We could say that the coupled term nullifies the magnetic term, since Eq.~(\ref{eq:pc}) has a negative contribution. Physically, this implies that the momentum diffusion constant of a magnetic dipole is dominated by time-dependent electric field fluctuation through $(1/c^2) \textbf{m}\times\partial_t \textbf{E}$ term instead of the magnetic field fluctuation through $\nabla(\textbf{m} \cdot \textbf{B})$ term in Eq. (\ref{eq:fx}). Hence, $\langle\Delta p\rangle^2/\Delta t$ will be dominated by the time-dependent fluctuations of the electric field. Note that the hierarchy is obtained under the assumptions of isotropic thermal bath and the long-wavelength approximation, where the wavelength of the electromagnetic radiation is larger than the size of the spatial superposition, which we will discuss below.  

Now, let us compute the finite-volume effect by considering the bulk diamagnetic property of the sphere with radius $a$ and $\alpha$ by using Clausius–Mossotti magnetic polarizability~\cite{stratton2007electromagnetic}:
\begin{equation}
\label{eq:polnano}
    \alpha =  \frac{a^3}{\mu_0} \frac{\mu - \mu_0}{\mu + 2\mu_0}= \frac{a^3}{\mu_0} \frac{\chi_v}{3 + \chi_v},
\end{equation}
Where $\mu$ and $\chi$ are the magnetic permeability and magnetic susceptibility, respectively. Both quantities are generally complex, with $\mu=\mu_R + i\mu_I$ and $\chi_v=\chi_{R}+i\chi_I$. Therefore, in this case, $\alpha_R(\omega)$, $\alpha_I(\omega)$ and $|\alpha(\omega)|^2$ can be written as
\begin{eqnarray}
    \alpha_R(\omega) &=& \frac{a^3}{\mu_0} \frac{\chi_R^2 +3\chi_R+\chi_I^2}{(3+\chi_R)^2+\chi_I^2}, \\
    \alpha_I(\omega) &=& \frac{a^3}{\mu_0} \frac{3\chi_I}{(3+\chi_R)^2+\chi_I^2}, \label{eq:imClauMos} \\
    |\alpha(\omega)|^2 &=& \frac{a^6}{\mu_0^2} \frac{(\chi_R^2 +3\chi_R+\chi_I^2)^2+ 9\chi_I^2}{[(3+\chi_R)^2+\chi_I^2]^2}. \label{eq:scaClauMos}
\end{eqnarray}
Note that $\alpha_I(\omega)$ is related to the extinction coefficient. However, if $\chi_I = 0$, the extinction would vanish, which is unphysical since there is still a contribution from scattering to extinction, given by $|\alpha(\omega)|^2$ (see Appendix \ref{sec:Extinction} and Ref. \cite{vanDeHulst1957}, Sec. 6.12). Therefore, the extinction coefficient $\alpha_I(\omega)$ should account for both scattering and absorption contributions, as in Eq. (\ref{eq:withabs}), while Eq. (\ref{eq:imClauMos}) represents only the absorption contribution \cite{vanDeHulst1957}. Consequently, they can be written as
\begin{eqnarray}
    \alpha_I^{\mathrm{abs}}(\omega) &=& \frac{a^3}{\mu_0} \frac{3\chi_I}{(3+\chi_R)^2+\chi_I^2} \label{eq:absClauMos} \\
    \alpha_I(\omega) &=& \frac{\mu_0}{6\pi} \left( \frac{\omega}{c} \right)^3\frac{a^6}{\mu_0^2} \frac{(\chi_R^2 +3\chi_R+\chi_I^2)^2+ 9\chi_I^2}{[(3+\chi_R)^2+\chi_I^2]^2} \beta + \frac{a^3}{\mu_0} \frac{3\chi_I}{(3+\chi_R)^2+\chi_I^2}. \label{eq:extClauMos}
\end{eqnarray}

The total momentum diffusion constant from Eq. (\ref{eq:genptot}) can be written in two terms.
\begin{equation} \label{eq:mdcwithabs}
    \frac{\langle \Delta p^2 \rangle}{\Delta t} = \frac{2 \mu_0^2 h^2}{9\pi^3 c^8} \int_0^{\infty} d\omega \omega^8 |\alpha(\omega)|^2 [n^2(\omega) + n(\omega)] + \frac{4 \mu_0 h^2 a^3}{3\pi^2 c^5 \beta} \int_0^{\infty} d\omega \omega^5 \alpha_I^{\mathrm{abs}}(\omega) [n^2(\omega) + n(\omega)].
\end{equation}
By substituting Eqs. (\ref{eq:scaClauMos}), (\ref{eq:absClauMos}) and (\ref{occupation number}), and
by assuming $\alpha(\omega),~\alpha_I^{\rm abs}(\omega)$ are constant, and by using the Clausius-Mossotti relation, Eq.~(\ref{eq:polnano}),
we would get
\begin{eqnarray}\label{eq:mdcfin}
 \frac{\langle \Delta p^2 \rangle}{\Delta t} &=& \frac{2 \hbar^2 a^6}{9\pi^3 c^8} \frac{(\chi_R^2 +3\chi_R+\chi_I^2)^2+ 9\chi_I^2}{[(3+\chi_R)^2+\chi_I^2]^2} \int_0^{\infty} d\omega \omega^8 \frac{e^{\hbar \omega /k_B T}}{(e^{\hbar \omega /k_B T} - 1)^2} + \frac{\hbar^2 a^3}{3\pi^2 c^5 \beta} \frac{3\chi_I}{(3+\chi_R)^2+\chi_I^2} \times \nonumber \\
    &&\int_0^{\infty} d\omega \omega^5 \frac{e^{\hbar \omega /k_B T}}{(e^{\hbar \omega /k_B T} - 1)^2}, \nonumber \\
    &=& \frac{2 \hbar^2 a^6 c}{9\pi^3} \frac{(\chi_R^2 +3\chi_R+\chi_I^2)^2+ 9\chi_I^2}{[(3+\chi_R)^2+\chi_I^2]^2} \left( \frac{k_B T}{\hbar c} \right)^9 \Gamma(9) \zeta(8) + \frac{\hbar^2 a^3 c}{3\pi^2 \beta} \frac{3\chi_I}{(3+\chi_R)^2+\chi_I^2} \times \\
    && \left( \frac{k_B T}{\hbar c} \right)^6 \Gamma(6) \zeta(5). \nonumber
\end{eqnarray}
For the case of pure nanodiamond, typically, the susceptibility is given by: $\chi_R= -2.2 \times 10^{-5}$ while the imaginary part $\mathrm{Im}(\chi_v) \approx 0$, see~\cite{richards_time-resolved_2025,yelisseyev_magnetic_2009,poklonski_magnetic_2023}. However, for an NV-centered nanodiamond, ${\rm Im}(\chi_v)\neq 0$, and it depends on the defect concentration, impurities and other defects and also the size of the nanodiamond, see~\cite{Afshar2025Magnetometry}, and it is not extremely well-known in a levitating setup for nanodiamonds. 

The expression $\langle \Delta p^2 \rangle/\Delta t$ is related to the momentum diffusion constant $D$, see \cite{Berg-Sorenson_1992,Balykin_1986,Dalibard_1985,Agarwal:1993}, where it fulfills this equation:
$    2D \Delta t  =  \lim_{\Delta t \rightarrow \infty} \langle \Delta p^2 \rangle$.
Therefore,  the momentum diffusion constant is $2D \equiv \langle \Delta p^2 \rangle/\Delta t$.
It was first observed in \cite{Sinha:2022snc} that the momentum diffusion constant 
is closely related to the scattering constant $\Lambda$ in the collisional
decoherence models; $2\Lambda = (1/\hbar^{2})\langle\Delta p^2\rangle/\Delta t$.
Their results matched the established decoherence rate computed by the scattering theory of blackbody radiation, see \cite{schlosshauer,RomeroIsart2011LargeQS}. The decoherence rate in the long-wavelength limit (i.e. the wavelength of the
environmental particle is much larger than the size of the
spatial superposition, which is given by $\gamma = \Lambda(\Delta x)^2$,
with the spatial superposition size given by $\Delta x$. Hence, our general result for the decoherence rate will be:
\begin{align}\label{eq:mdcdecoherence}
\gamma =&\Big[\frac{a^6 c}{9\pi^3} \frac{(\chi_R^2 +3\chi_R+\chi_I^2)^2+ 9\chi_I^2}{[(3+\chi_R)^2+\chi_I^2]^2} \left( \frac{k_B T}{\hbar c} \right)^9 \Gamma(9) \zeta(8) + \frac{a^3 c}{6\pi^2 \beta} \frac{3\chi_I}{(3+\chi_R)^2+\chi_I^2} \left( \frac{k_B T}{\hbar c} \right)^6 \nonumber \\
& \times \Gamma(6) \zeta(5) \Big](\Delta x)^2,
\end{align}
which includes both the real and imaginary part of the $\alpha$, and the magnetic susceptibility $\chi_v=\chi_{R}+i\chi_{I}$. Note that the final expression of decoherence and the momentum diffusion constant in Eq. (\ref{eq:mdcfin}) are valid in the long-wavelength limit and isotropic thermal bath. In the case, when $\chi_{I}\approx 0$, the first term dominates, and the corresponding momentum diffusion constants can be found in the Appendix D.


\section{Interference and Decoherence}
\label{sec:decoherence}

Typically, in the QGEM experiment~\cite{Bose:2017nin}, two nanodiamonds interact via background electromagnetic and gravitational interactions, as well. Hence, it is important to understand how the two levitated nanodiamonds will interact with the external electromagnetic field and their fluctuations. Here, we will focus only on the background electromagnetic fluctuations and not specifically on the chip, which we had already considered in \cite{Elahi:2024dbb, Xiang:2026kwd}.
Let us consider the interaction between the two magnetic dipoles, $m_1$ and $m_2$, located at $x_1$ and $x_2$. The interaction Hamiltonian of these dipoles with the magnetic field is given by~\footnote{A similar analysis for the point dipole in the presence of the electric field has been analysed in \cite{Sinha:2022snc}. Here, we compute the magnetic dipole case and compare the results. }:
\begin{equation} \label{eq:H_I}
    H_I= - \textbf{m}_1 \cdot \textbf{B}(\textbf{x}_1,t) - \textbf{m}_2 \cdot \textbf{B}(\textbf{x}_2,t).
\end{equation}
The quantisation of the magnetic field is $\textbf{B}$ from Eq. (\ref{eq:B}). On the other hand, the free-field Hamiltonian of the photon (electromagnetic field) can be written as:
\begin{equation} \label{eq:H_F}
    H_F = \hbar \omega \left(a_{\textbf{k} \lambda}^{\dagger} a_{\textbf{k} \lambda} + \frac{1}{2}\right)
\end{equation}
In this section, we determine the decoherence rate from the interference of the scattered photons by the two localised magnetic dipoles. The interference of the scattered photons refers to the quantum overlap in the wavefunctions of thermal photons scattered from the two dipole positions; when their respective spatial superposition, which is assumed here to be the same, $\Delta x$, is small compared to the distance between their centre-of-mass. The photons are indistinguishable, allowing constructive interference in the emission rate, but as the $\Delta x$ become larger, they become distinguishable and carry away "which-path" information, reducing interference and causing decoherence \cite{schlosshauer}. In the Heisenberg picture, the momentum-diffusion constant is an ensemble average over momentum kicks. Therefore, we start from the Heisenberg picture equation of motion, since decoherence and the momentum diffusion constant are related. The equation of motion is given by:
\begin{equation}
\label{eq:heis}
    i \hbar \frac{d}{dt'}{a}_{\textbf{k} \lambda}= [a_{\textbf{k} \lambda}, H_F] + [a_{\textbf{k} \lambda}, H_I].
\end{equation}
Using Eq. (\ref{eq:H_F}) on the first commutators on Eq. [\ref{eq:heis}] could be written as
\begin{eqnarray}
    [a_{\textbf{k} \lambda}, H_F] =& \hbar \omega a_{ \textbf{k} \lambda}(t').
\end{eqnarray}
Similarly, using Eqs. [\ref{eq:B}] and[\ref{eq:H_I}] into the second term of Eq. [\ref{eq:heis}], we get
\begin{equation}
    [a_{\textbf{k} \lambda}, H_I] = i \sum_{i=1}^2 \left( \frac{\hbar}{2 \varepsilon_0 \omega V} \right)^{1/2} e^{-i \textbf{k} \cdot \textbf{x}} [\textbf{m}_i (t') \cdot (\textbf{k} \times \textbf{e}_{\textbf{k} \lambda})].
\end{equation}
By using these commutators and substituting them into Eq. (\ref{eq:heis})
\begin{eqnarray}
\label{eq:solheis}
    a_{\textbf{k} \lambda} (t) &=& a_{\textbf{k} \lambda} (0) e^{-i \omega t} + \sum_{i=1}^2 \left( \frac{1}{2 \hbar \varepsilon_0 \omega V} \right)^{1/2} e^{-i \textbf{k} \cdot \textbf{x}} \int_0^t dt' e^{i \omega (t'-t)} [\textbf{m}_i (t') \cdot (\textbf{k} \times \textbf{e}_{\textbf{k} \lambda})], \\
    &=& a_{\textbf{k} \lambda} (0) e^{-i \omega t} + a_{\textbf{k} \lambda}^{\mathrm{(int)}}(t)\,, \nonumber
\end{eqnarray}
where the first term is the free-field term and the second term is the dipole-field interaction term. Although the second term contains an integral over $t'$, its overall time dependence remains proportional to $e^{-i \omega t}$, and therefore it contributes to the annihilation operator, $a_{{\rm k}\lambda}^{\rm int}(t)$, in the Heisenberg picture.
The rate of the average photon number radiated by the two magnetic dipoles \cite{milonni2019introduction} induced at $\textbf{x}_1$ and $\textbf{x}_2$ is given by the rate of change of the occupation number of the photons after the interaction, given by $a_{\textbf{k} \lambda}^{\mathrm{(int)} \dagger}(t) a_{\textbf{k} \lambda}^{\mathrm{(int)}}(t)$.
\begin{equation}
R= \frac{d}{dt}(\langle a_{\textbf{k} \lambda}^{\mathrm{(int)} \dagger}(t) a_{\textbf{k} \lambda}^{\mathrm{(int)}}(t)\rangle).     
\end{equation}
Using eq. \ref{eq:solheis}, we could write $R$ as 
\begin{equation}
\label{eq:R}
    R = \frac{d}{dt} \sum_{\textbf{k} \lambda}\langle (a_{\textbf{k} \lambda}^{\dagger} (t) - a_{\textbf{k} \lambda}^{\dagger} (0) e^{i \omega t})(a_{\textbf{k} \lambda} (t) - a_{\textbf{k} \lambda} (0) e^{-i \omega t}) \rangle,
\end{equation}
or
\begin{align} \label{eq:Rsub}
    R &= \frac{d}{dt}\sum_{i,j=1}^2 \sum_{\textbf{k} \lambda} \left( \frac{1}{2 \hbar \varepsilon_0 \omega V} \right) e^{i \textbf{k} \cdot (\textbf{x}_i - \textbf{x}_j)} \int_0^t dt'' \int_0^t dt' e^{i \omega (t'-t'')} \langle [\textbf{m}_i^{\dagger}(t'') \cdot(\textbf{k} \times \textbf{e}_{\textbf{k} \lambda})] \nonumber \\
    &~~~~~ [\textbf{m}_j(t') \cdot(\textbf{k} \times \textbf{e}_{\textbf{k} \lambda})] \rangle, \\
    &= \frac{1}{ \hbar \varepsilon_0 V}  \sum_{i,j=1}^2 \sum_{\textbf{k}} \frac{e^{i \textbf{k} \cdot (\textbf{x}_i - \textbf{x}_j)}}{\omega} \mathrm{Re}\left[\int_0^t dt' e^{i \omega (t'-t)} \sum_{\lambda} \langle [\textbf{m}_i^{\dagger}(t) \cdot(\textbf{k} \times \textbf{e}_{\textbf{k} \lambda})] [\textbf{m}_j(t') \cdot(\textbf{k} \times \textbf{e}_{\textbf{k} \lambda})] \rangle \right]. \nonumber
\end{align}
Let us inspect the time integration, defined as $b_{ij}$, where $i,j=1,2$.
\begin{equation}
\label{eq:xij}
    b_{ij} = \int_0^t dt' e^{i \omega (t'-t)} \sum_{\lambda} \langle [\textbf{m}_i^{\dagger}(t) \cdot(\textbf{k} \times \textbf{e}_{\textbf{k} \lambda})] [\textbf{m}_j(t') \cdot(\textbf{k} \times \textbf{e}_{\textbf{k} \lambda})] \rangle
\end{equation}
The magnetic dipole can be written has the same expression as Eq. (\ref{eq:mj}) in vector form. Hence, the dot product of the magnetic dipole moment can be written as
\begin{eqnarray}
    \textbf{m}_i(t) \cdot (\textbf{k} \times \textbf{e}_{\textbf{k} \lambda}) &=& i \sum_{\textbf{k}' \lambda'} \left( \frac{\hbar}{2 \varepsilon_0 \omega' V} \right)^{1/2} \alpha(\omega') [a_{\textbf{k}' \lambda'}(0) e^{-i \omega' t} e^{i \textbf{k} \cdot \textbf{x}_i} - a_{\textbf{k}' \lambda'}^{\dagger}(0) e^{i \omega' t} e^{-i \textbf{k} \cdot \textbf{x}_i}] \nonumber \\
    && [(\textbf{k} \cdot \textbf{k}')(\textbf{e}_{\textbf{k} \lambda} \cdot \textbf{e}_{\textbf{k}' \lambda'})-(\textbf{k} \cdot \textbf{e}_{\textbf{k}' \lambda'})(\textbf{k}' \cdot \textbf{e}_{\textbf{k} \lambda})]
\end{eqnarray}
Here we used $(\textbf{k}' \times \textbf{e}_{\textbf{k}' \lambda'}) \cdot (\textbf{k} \times \textbf{e}_{\textbf{k} \lambda}) = (\textbf{k} \cdot \textbf{k}')(\textbf{e}_{\textbf{k} \lambda} \cdot \textbf{e}_{\textbf{k}' \lambda'})-(\textbf{k} \cdot \textbf{e}_{\textbf{k}' \lambda'})(\textbf{k}' \cdot \textbf{e}_{\textbf{k} \lambda})$. The equation (\ref{eq:xij}) can then be written as
\begin{eqnarray}
\label{eq:xijaf}
    b_{ij} &=& \sum_{\textbf{k'} \lambda'} \sum_{\lambda} \left( \frac{\hbar}{2 \varepsilon_0 \omega' V} \right) |\alpha(\omega')|^2 [\langle a_{\textbf{k}' \lambda'}^{\dagger}(0) a_{\textbf{k}' \lambda'}(0) \rangle e^{-i \textbf{k}' \cdot (x_i - x_j)} \int_0^t dt' e^{i(\omega - \omega')(t'-t)} \nonumber \\ 
    &&+ \langle a_{\textbf{k}' \lambda'}(0) a_{\textbf{k}' \lambda'}^{\dagger}(0) \rangle e^{i \textbf{k}' \cdot (x_i - x_j)} \int_0^t dt' e^{i(\omega + \omega')(t'-t)}] [(\textbf{k} \cdot \textbf{k}')(\textbf{e}_{\textbf{k} \lambda} \cdot \textbf{e}_{\textbf{k}' \lambda'}) \nonumber \\
    &&-(\textbf{k} \cdot \textbf{e}_{\textbf{k}' \lambda'})(\textbf{k}' \cdot \textbf{e}_{\textbf{k} \lambda})]^2.
\end{eqnarray}
The second term will be neglected since it is a non-resonant term, and by using 
\begin{eqnarray}
    \langle a_{\textbf{k}' \lambda'}^{\dagger}(0) a_{\textbf{k} \lambda} (0) \rangle &=& n_{\textbf{k}'} \delta^3_{k'k} \delta_{\lambda' \lambda},
\end{eqnarray}
then the Eq. (\ref{eq:xijaf}) can be written as
\begin{eqnarray}
    b_{ij} &=& \frac{\hbar}{2 \varepsilon_0 V} \frac{V}{8\pi^3} \int d^3 k' \frac{|\alpha(\omega')|^2}{\omega'}   n_{\textbf{k}'} e^{-i \textbf{k}' \cdot (\textbf{x}_i - \textbf{x}_j)} \int_0^t dt' e^{i(\omega- \omega')(t' - t)} \nonumber \\
    &&[(\textbf{k} \cdot \textbf{k}')(\textbf{e}_{\textbf{k} \lambda} \cdot \textbf{e}_{\textbf{k}' \lambda'})-(\textbf{k} \cdot \textbf{e}_{\textbf{k}' \lambda'})(\textbf{k}' \cdot \textbf{e}_{\textbf{k} \lambda})]^2.
\end{eqnarray}
Therefore, we could write
\begin{eqnarray}
\label{eq:R1}
    R_{ij} &=& \frac{1}{2 \varepsilon_0^2 V^2} \frac{V^2}{64\pi^6} \sum_{i,j=1}^2 \int \frac{d^3 k}{\omega} \int \frac{d^3 k'}{\omega'} |\alpha(\omega')|^2 n_k [(\textbf{k} \cdot \textbf{k}')(\textbf{e}_{\textbf{k} \lambda} \cdot \textbf{e}_{\textbf{k}' \lambda'})-(\textbf{k} \cdot \textbf{e}_{\textbf{k}' \lambda'})(\textbf{k}' \cdot \textbf{e}_{\textbf{k} \lambda})]^2 \nonumber \\
    && \times e^{i(\textbf{k}-\textbf{k}')\cdot (\textbf{x}_i - \textbf{x}_j)} \int_0^t dt' \cos{[(\omega - \omega')(t'-t)]}.
\end{eqnarray}
For $ t \rightarrow \infty$,
\begin{equation}
    \int_0^t dt' \cos{[(\omega - \omega')(t'-t)]} = \pi \delta(\omega - \omega')= \frac{\pi}{c} \delta(k - k').
\end{equation}
Therefore, Eq. (\ref{eq:R1}) becomes
\begin{eqnarray}
\label{eq:RRij}
    R_{ij} &=& \frac{1}{128 \pi^6 \varepsilon_0^2} \sum_{i,j=1}^2 \int \frac{d^3 k}{kc} \int \frac{d^3 k'}{k'c} n_{k'} e^{i(\textbf{k}-\textbf{k}')\cdot (\textbf{x}_i - \textbf{x}_j)} (kk')^2 (1+\cos^2 \theta) \delta(k-k') \frac{\pi}{c}, \nonumber \\
    &=& \frac{\mu_0^2 c}{128 \pi^5} \sum_{i,j=1}^2 \int dk k^6 |\alpha(\omega)|^2 n_k \int d\Omega_{\hat{\textbf{k}}} \int d\Omega_{\hat{\textbf{k}}'} (1+\cos^2 \theta) e^{ik(\hat{\textbf{k}}-\hat{\textbf{k}}') \cdot (\textbf{x}_i - \textbf{x}_j)}, 
\end{eqnarray}
Where $\hat{\textbf{k}}$ is the unit vector of direction of $\textbf{k}$, $\theta$ is angle between $\hat{\textbf{k}}$ and $\hat{\textbf{k}}'$, and $d\Omega_{\hat{\textbf{k}}}$ is the angular element of $\hat{\textbf{k}}$. In the previous section, we showed the relation between the polarizability $|\alpha(\omega)|^2$ and its imaginary part $\alpha_I$ through the extinction and scattering cross sections of a magnetic dipole in the Rayleigh scattering limit \cite{Neuringer1966,stratton2007electromagnetic} on Eq. (\ref{eq:noabs}). Since Eq. (\ref{eq:noabs}) accounts only for elastic scattering, the absorption effect is not included. Therefore, the absorption contribution is added in Eq. (\ref{eq:withabs}). Following the same procedure as Eqs. (\ref{eq:noabs})-(\ref{eq:withabs}), we obtain
\begin{eqnarray}\label{eq:Rabs}
    R_{ij} &=& \frac{3\mu_0 c}{64 \pi^4} \sum_{i,j=1}^2 \int dk k^3 \left[ \frac{\mu_0}{6\pi} k^3 |\alpha(\omega)|^2 + \frac{1}{\beta}\alpha_I^{\mathrm{abs}}(\omega)  \right] n_k \int d\Omega_{\hat{\textbf{k}}} \int d\Omega_{\hat{\textbf{k}}'} (1+\cos^2 \theta) \times \nonumber \\
    && e^{ik(\hat{\textbf{k}}-\hat{\textbf{k}}') \cdot (\textbf{x}_i - \textbf{x}_j)}.
\end{eqnarray}

By using $n_{k'}=(e^{\hbar k c/k_BT}-1)^{-1}$, the Eq. (\ref{eq:Rabs}) can be written as
\begin{equation}
\label{eq:R12}
    R_{12} = R_{21} = \frac{3 \mu_0 c}{64 \pi^4}  \int dk \frac{k^3 \left[ \frac{\mu_0}{6\pi} k^3 |\alpha(\omega)|^2 + \frac{1}{\beta}\alpha_I^{\mathrm{abs}}(\omega)  \right]}{e^{\hbar k c/k_B T}-1} \int d\Omega_{\hat{\textbf{k}}} \int d\Omega_{\hat{\textbf{k}}'} (1+\cos^2 \theta) e^{ik(\hat{\textbf{k}}-\hat{\textbf{k}}') \cdot (\textbf{x}_1 - \textbf{x}_2)} .
\end{equation}
Similarly,
\begin{equation}
\label{eq:R11}
    R_{11} = R_{22} = \frac{3\mu_0 c}{64 \pi^4} \int dk \frac{k^3 \left[ \frac{\mu_0}{6\pi} k^3 |\alpha(\omega)|^2 + \frac{1}{\beta}\alpha_I^{\mathrm{abs}}(\omega)  \right]}{e^{\hbar k c/k_bT}-1} \int d\Omega_{\hat{\textbf{k}}} \int d\Omega_{\hat{\textbf{k}}'}  (1+\cos^2{\theta})).
\end{equation}
Unlike $R_{11}$, the rate $R_{12}$ is a photon scattering rate that involves interference between the magnetic fields of the two dipoles. Consequently, the difference $R_{11} - R_{12}$ is a measure of the rate at which this interference decreases as the distance between the dipoles increases. We could say it describes the loss of spatial coherence between the dipoles, or decoherence factor.\footnote{Generally, the decoherence factor should be a combination of $\frac{1}{2}(R_{11} + R_{22} - R_{12} - R_{21})$, where $1/2$ is a prefactor on Lindblad coefficient $\kappa$, see chapter 4 Ref. \cite{schlosshauer}. Since $R_{12}=R_{21}$ and $R_{11}=R_{22}$ are symmetry, it reduces to $(R_{11} - R_{12})$.
On the other hand, the differences between $R_{11} - R_{22}$ and $R_{12} - R_{21}$ do not have contribution to the decoherence. Furthermore, once we assume $m_1=m_2$, we can treat the scenario as a dipole being a spatial superposition, i.e. $\Delta x=x_2-x_1$.} 
Hence, we could write them as
\begin{eqnarray}
\label{eq:RSI}
    F(\textbf{x}_1 - \textbf{x}_2) &=& R_{11} - R_{12} \nonumber \\
    &=& \frac{3 \mu_0 c}{64 \pi^4} \int dk \frac{k^3 \left[ \frac{\mu_0}{6\pi} k^3 |\alpha(\omega)|^2 + \frac{1}{\beta}\alpha_I^{\mathrm{abs}}(\omega)  \right]}{e^{\hbar k c/k_B T}-1} \int d\Omega_{\hat{\textbf{k}}} \int d\Omega_{\hat{\textbf{k}}'} (1+\cos^2 \theta) \times  \\
    && (1 - e^{ik(\hat{\textbf{k}}-\hat{\textbf{k}}') \cdot (\textbf{x}_1 - \textbf{x}_2)}). \nonumber
\end{eqnarray}
This decoherence factor is for the general magnetic nanoparticles, where we have included the absorption term. In the long-wavelength limit, $k<< (\textbf{x}_1-\textbf{x}_2)$ \footnote{For creating a quantum superposition with masses $m=10^{-14}$ kg and $\Delta x \sim {\cal O}(10) {\rm \mu m}$, see Ref. \cite{SchutMazumdar2025}, the  long-wavelength regime entails that the wavelength of the electromagnetic radiation must be larger than $\lambda \gg 10 {\rm \mu m}$, since $k=2\pi/\lambda$.}, we use the approximation.
\begin{equation}
    1 - e^{ik(\hat{\textbf{k}}-\hat{\textbf{k}}') \cdot (\textbf{x}_1 - \textbf{x}_2)} \approx -ik(\hat{\textbf{k}}-\hat{\textbf{k}}') \cdot (\textbf{x}_1 - \textbf{x}_2) + \frac{1}{2}k^2 [(\hat{\textbf{k}}-\hat{\textbf{k}}') \cdot (\textbf{x}_1 - \textbf{x}_2)]^2.
\end{equation}
Where the first term doesn't contribute to the integral because the dot product of the first term is an odd function. When integrated over all directions of $\hat{\textbf{k}}$ and $\hat{\textbf{k}}'$, its product with an even function such as $(1+ \cos^2 \theta)$ vanishes due to symmetry\footnote{The same argument explains why $R_{12}=R_{21}$ in Eq. (\ref{eq:R12}).}. Therefore
\begin{eqnarray}
\label{eq:defactor1}
    F(\textbf{x}_1-\textbf{x}_2) &=& \frac{3\mu_0 c}{64\pi^4} \int dk \frac{k^3 \left[ \frac{\mu_0}{6\pi} k^3 |\alpha(\omega)|^2 + \frac{1}{\beta}\alpha_I^{\mathrm{abs}}(\omega)  \right]}{e^{\hbar k c/k_B T}-1} \int d\Omega_{\hat{\textbf{k}}} \int d\Omega_{\hat{\textbf{k}}'}  (1+\cos^2 \theta) \\
    && \times \frac{k^2}{2} [(\hat{\textbf{k}}-\hat{\textbf{k}}') \cdot (\textbf{x}_1 - \textbf{x}_2)]^2. \nonumber
\end{eqnarray}
Moreover, the dot product can be computed as an average over all directions of $\textbf{x}_1 - \textbf{x}_2$
\begin{eqnarray}
    (x_1-x_2)^2\frac{1}{3} \sum_{i=x,y,z} \left[(\hat{\textbf{n}} - \hat{\textbf{n}}') \cdot \hat{i} \right]^2 &=& \frac{2}{3} (x_1-x_2)^2 (1 - \hat{\textbf{n}} \cdot \hat{\textbf{n}}') \nonumber \\
    &=& \frac{2}{3} (x_1-x_2)^2(1- \cos \theta).
\end{eqnarray}
Hence Eq. (\ref{eq:defactor1}) can be written as
\begin{eqnarray}\label{eq:defactor2}
    F(\textbf{x}_1- \textbf{x}_2) &=& \frac{3 \mu_0 c}{64 \pi^4} \int dk \frac{k^3 \left[ \frac{\mu_0}{6\pi} k^3 |\alpha(\omega)|^2 + \frac{1}{\beta}\alpha_I^{\mathrm{abs}}(\omega)  \right]}{e^{\hbar k c/k_B T}-1} \int d\Omega_{\hat{\textbf{k}}} \int d\Omega_{\hat{\textbf{k}}'} (1+\cos^2 \theta) \nonumber \\
    && \times \frac{1}{3}k^2 (x_1-x_2)^2(1- \cos \theta). 
\end{eqnarray}
Using polarizability of a nanodiamagnetic sphere in Eq. (\ref{eq:polnano}) and the angular integration part as
\begin{equation}
    \int d\Omega_{\hat{\textbf{k}}} \int d\Omega_{\hat{\textbf{k}}'} (1+\cos^2{\theta}) (1- \cos{\theta}) = \frac{64\pi^2}{3}.
\end{equation}
The decoherence factor becomes by assuming $\alpha(\omega)$ is constant, and by using the $\alpha(\omega)$ Clausius-Mossotti relation, Eqs.~\ref{eq:scaClauMos} and ~\ref{eq:absClauMos}:
\begin{eqnarray} \label{eq:mag}
    F(\textbf{x}_1- \textbf{x}_2) &=& (x_1 - x_2)^2 \frac{a^6 c}{18 \pi^3} \frac{(\chi_R^2 +3\chi_R+\chi_I^2)^2+ 9\chi_I^2}{[(3+\chi_R)^2+\chi_I^2]^2} \int_0^{\infty} dk \frac{k^8}{e^{\hbar k c / k_B T}-1} + \nonumber \\
    && (x_1 - x_2)^2 \frac{a^3 c}{3\pi^2 \beta} \frac{3\chi_I}{(3+\chi_R)^2+\chi_I^2} \int_0^{\infty} dk \frac{k^5}{e^{\hbar k c / k_B T}-1}, \nonumber \\
    &=& (x_1 - x_2)^2 \frac{a^6 c}{18 \pi^3} \frac{(\chi_R^2 +3\chi_R+\chi_I^2)^2+ 9\chi_I^2}{[(3+\chi_R)^2+\chi_I^2]^2} \left( \frac{k_B T}{\hbar c} \right)^9 \zeta(9) \Gamma(9) + \nonumber \\
    && (x_1 - x_2)^2 \frac{a^3 c}{3\pi^2 \beta} \frac{3\chi_I}{(3+\chi_R)^2+\chi_I^2} \left( \frac{k_B T}{\hbar c} \right)^6 \Gamma(6)\zeta(6). 
\end{eqnarray}
Let us compare it with the decoherence factor of a dielectric sphere in an electromagnetic field, which is given by \cite{Sinha:2022snc}:
\begin{equation} \label{eq:elec}
    F_{E}(\textbf{x}_1- \textbf{x}_2) = (x_1-x_2)^2 \frac{8a^6 c}{9 \pi}  \left| \frac{\varepsilon - 1}{\varepsilon + 2 } \right|^2 \left( \frac{k_B T}{\hbar c} \right)^9 \zeta(9) \Gamma(9).
\end{equation}
Where $\varepsilon$ is the relative permittivity $\varepsilon=\varepsilon_m/\varepsilon_0$~\footnote{In principle, permittivity also has a real and imaginary component, $\varepsilon=\varepsilon_R + \varepsilon_I$. However, we only consider the real part of $\varepsilon$.}. By assuming that $\chi_I \approx 0$, we can directly compare our results of  Eqs.~(\ref{eq:defactor2},\ref{eq:mag}) to that of the dielectric sphere case. Hence, we will get a relation between the decoherence rate of the dielectric sphere and the diamagnetic sphere, as:
\begin{equation}\label{ratio-2diamag}
    \frac{F_{B}(\textbf{x}_1- \textbf{x}_2)}{F_{E}(\textbf{x}_1- \textbf{x}_2)} = \frac{1}{16\pi^2}  \frac{\chi_R^2}{(3+\chi_R)^2} \left| \frac{\varepsilon + 2}{\varepsilon - 1} \right|^{2}  .
\end{equation}
We will discuss the physical interpretation of this result in the discussion section.

\section{Drag Force}
\label{sec:dragforce}

In this last section, we will discuss the 
Einstein and Hopf effect~\cite{Einstein:1910a,Einstein:1910b} on a moving nanoparticle with a magnetic dipole. Einstein and Hopf showed that a moving polarizable particle would experience a drag force $F=-\xi mv$ with velocity $v$, where $v<<c$, with respect to a frame where there are isotropic fields. Here, we wish to compute the drag force on magnetic dipole moments in a magnetic field in a relativistic form first and then take the non-relativistic limit. This computation is essential for any levitated experiment, especially levitating the nanoparticle via magnetic field. Any fluctuations in the external magnetic field will lead to a drag force on the nanoparticle. Again, utilising that the external magnetic field and it's gradient is such that we can use  a  point like magnetic dipole moment approximation, see~\cite{Elahi:2024dbb}. In our case, let's assume magnetic dipole moments move along the x-axis in a laboratory frame $S$. The Lorentz transformation of a particle in a laboratory frame and the particle's frame $S'$
\begin{eqnarray}
    x' = \gamma(x-vt),~~~
    t' = \gamma \left(t-\frac{vx}{c^2} \right),~~~
    p_x' = \gamma \left( p_x - \frac{v}{c^2} \varepsilon \right),~~~
    \varepsilon = \gamma ( \varepsilon - v p_x).
\end{eqnarray}
where $dx/dt = v$, $\gamma=(1-v^2/c^2)^{-1/2}$, and $\varepsilon$ is the energy. These transformations affect the force $F_x$ in $S$ frame and $F_x'$ in $S'$ 
\begin{eqnarray} \label{eq:Ftot}
    F_x = \frac{dp_x}{dt} = \frac{d}{dt'} \left( p_x' + \frac{v}{c^2} \varepsilon' \right) \frac{dt'}{dt}
    = \frac{dp_x'}{dt'} + \frac{v}{c^2} \frac{d\varepsilon'}{dt'}
    = F_x' + \frac{v}{c^2} \frac{d\varepsilon'}{dt'}.
\end{eqnarray}
The force acting on the dipole is divided into two contributions: $F_{ind}$, the force from the dipole induced by fluctuating fields, and $F_{dd}$ force from the dipole's emitted radiation field and fluctuations\footnote{Our focus will be on the $F_{ind}$ for now.}. Our focus is on calculating $F_{ind}$, which is computed by starting from $F_x'$ and $d\varepsilon'/dt'$. However, $F_{ind}$ doesn't have a contribution from $d\varepsilon'/dt'$ since $d\varepsilon'/dt' = d/dt(\textbf{F}\cdot \textbf{u})$, and $\textbf{u}=0$ in the particle's rest frame. Hence, we will only compute the first term. The magnetic field could be written in the dipoles' rest frame as
\begin{eqnarray}
    B_x' = B_x,~~~~
    B_y' = \gamma \left(B_y + \frac{v}{c} E_z \right),~~~~
    B_z' = \gamma \left(B_z - \frac{v}{c} E_y \right).
\end{eqnarray}
Hence, the force acting on the dipole moment in presence of the external magnetic field in the $x$-axis can be written as:
\begin{equation}
    F_x = m_x \frac{\partial B_x}{\partial x} + m_x \frac{\partial B_y}{\partial x} + m_x \frac{\partial B_z}{\partial x}.
\end{equation}
The magnetic field in a single field mode polarized along the $x$-direction can be written as
\begin{eqnarray}
    B_x' &=& B_x = i \left( \frac{\hbar}{2 \varepsilon_0 \omega V} \right)^{1/2} \left[ a_{\textbf{k} \lambda}e^{-i(\omega t - \textbf{k} \cdot \textbf{x})} - a_{\textbf{k} \lambda}^{\dagger} e^{i(\omega t - \textbf{k} \cdot \textbf{x})} \right] (\textbf{k} \times \textbf{e}_{\textbf{k} \lambda})_x, \nonumber \\
    &=& i \left( \frac{\hbar}{2 \varepsilon_0 \omega V} \right)^{1/2} \left[ a_{\textbf{k} \lambda}e^{-i(\omega' t' - \textbf{k}' \cdot \textbf{x}')} - a_{\textbf{k} \lambda}^{\dagger} e^{i(\omega' t' - \textbf{k}' \cdot \textbf{x}')} \right] (\textbf{k} \times \textbf{e}_{\textbf{k} \lambda})_x.
\end{eqnarray}
where we have used that
\begin{eqnarray}
    \omega t - \textbf{k} \cdot \textbf{x} = \omega' t' - \textbf{k}' \cdot \textbf{x}',~~~
    \omega' = \gamma(\omega - v k_x),~~~
    k'_x = \gamma \left(k_x - \frac{v}{c^2} \omega'\right). 
\end{eqnarray}
The induced magnetic dipole in a frame of the reference of a nanoparticle is given by
\begin{equation}
    m_x' = i \left( \frac{\hbar}{2 \varepsilon_0 \omega V} \right)^{1/2} \left[ \alpha(\omega') a_{\textbf{k} \lambda}e^{-i(\omega' t' - \textbf{k}' \cdot \textbf{x}')} - \alpha^*(\omega') a_{\textbf{k} \lambda}^{\dagger} e^{i(\omega' t' - \textbf{k}' \cdot \textbf{x}')} \right] (\textbf{k} \times \textbf{e}_{\textbf{k} \lambda})_x.
\end{equation}
Hence, the force in the $S'$, in the nanoparticle's, frame can be written as
\begin{eqnarray}
    [F_{ind}]_{xx} &=& \mathrm{Re}\left \langle m'_x \frac{\partial B_x'}{\partial t'} \right \rangle, \nonumber \\
    &=& \mathrm{Re} \left[ -i\frac{\hbar}{2 \varepsilon_0 \omega V} k_x' \{ \alpha(\omega')( n(\omega) + 1) - \alpha^*(\omega') n(\omega) \}  \right] (\textbf{k} \times \textbf{e}_{\textbf{k} \lambda})_x^2, \nonumber \\
    &=& \frac{\hbar}{2\varepsilon_0 \omega V} k'_x \alpha_I(\omega') [2n(\omega)+1] (\textbf{k} \times \textbf{e}_{\textbf{k} \lambda})_x^2
\end{eqnarray} 
Note that the $\alpha_R(\omega')$ cancels out, leaving behind the imaginary contribution, i.e. $\alpha_I(\omega')$.
Here, the expectation value of photon annihilation and creation operators have been used, and $\alpha_I(\omega')$ is the imaginary part of polarizability, which is related to the extinction cross section including the scattering and absorption contribution, as given in Eq. (\ref{eq:withabs}). The force from all modes can be written as
\begin{eqnarray}
    [F_{ind}]_{xx} &=& \frac{V}{8\pi^3} \int d^3 k \frac{\hbar}{\varepsilon_0 \omega V} k'_x \alpha_I(\omega') \left[n(\omega)+\frac{1}{2}\right] \sum_{\lambda=1}^2 (\textbf{k} \times \textbf{e}_{\textbf{k} \lambda})_x^2, \nonumber \\
    &=& \frac{\hbar}{\pi^3 \varepsilon_0} \frac{V}{8\pi^3} \int d^3 k \frac{(k^2 - k_x^2)}{\omega} k'_x \alpha_I(\omega') \left[n(\omega)+\frac{1}{2}\right]. \nonumber \\
    &=& \frac{\hbar}{8 \pi^3 \epsilon_o c} \int d^3 k' k' k_x' \left( 1 - \frac{k_x'^2}{k'^2}   \right) \alpha_I(\omega') \left[n (\gamma(\omega' + vk_x'))+ \frac{1}{2}\right]
\end{eqnarray}
Where we use Jacobian and $\omega'$ as
\begin{eqnarray}
    d^3 k = \gamma \left( 1 + \frac{v}{c} \frac{k'^2_x}{k'^2} \right) d^3 k',~~~~~
    \omega = \gamma(\omega' + v k_x').
\end{eqnarray}
Furthermore, by using the same computation, we could get
\begin{eqnarray}
    [F_{ind}]_{xy} &=& \frac{\hbar}{8\pi^3 \epsilon_o c} \int d^3 k' k' k_x' \left( 1 - \frac{k_y'^2}{k'^2}   \right) \alpha_I(\omega') \left[n (\gamma(\omega' + vk_x'))+ \frac{1}{2}\right], \\
    \left[F_{ind} \right]_{xz} &=& \frac{\hbar}{8 \pi^3 \epsilon_o c} \int d^3 k' k' k_x' \left( 1 - \frac{k_z'^2}{k'^2}   \right) \alpha_I(\omega') \left[n (\gamma(\omega' + vk_x'))+ \frac{1}{2}\right].
\end{eqnarray}
Hence, the final expression of the total force in $x$-direction
\begin{eqnarray} \label{eq:FindM}
    F_{ind} &=& [F_{ind}']_{xx} + [F_{ind}']_{xy} + [F_{ind}']_{xz}
    = \frac{\hbar}{4 \pi^3 \epsilon_o c} \int d^3 k' k' k_x' \alpha_I(\omega') \left[n (\gamma(\omega' + vk_x'))+ \frac{1}{2}\right]. \\
    &=& \mu_0 \frac{\hbar c}{4 \pi^3} \int d^3 k k k_x \alpha_I(\omega) n (\gamma(\omega' + vk_x')). \nonumber
\end{eqnarray}
Where we have omitted the prime because the two frames are approximately similar in the non-relativistic limit $(v<<c)$, and factor of $1/2$ disappeared after angular integration of $d\Omega_k$. Here, we take the Taylor expansion of order $v/c$.
\begin{equation}
    n (\gamma(\omega + vk_x)) \approx n(\omega) + vk_x \frac{dn}{d\omega}.
\end{equation}
Substitute Eq. above to Eq. (\ref{eq:FindM}) and do the angular part of the integration, we get
\begin{equation} \label{eq:Findtot}
    F_{ind} = \mu_0 \frac{\hbar}{3 \pi^2 c^5} v \int_0^{\infty}d\omega \omega^5 \alpha_I(\omega) \frac{dn}{d\omega} \equiv -\xi mv.
\end{equation}
Thus, we also obtained, in this non-relativistic limit, the drag force on magnetic dipoles in a magnetic field in a laboratory frame. Simplifying (\ref{eq:Findtot}), one obtain
\begin{equation}\label{eq:thermaldrag}
    \xi = \mu_0 \frac{\hbar}{3 \pi^2 m c^5} \int_0^{\infty}d\omega \omega^5 \alpha_I(\omega) \frac{dn}{d\omega}.
\end{equation}
Therefore, substituting $\alpha_I$ into Eq. (\ref{eq:withabs}) where it represents the extinction coefficient including scattering and absorption contributions, the thermal drag for a nanodiamagnetic sphere can be written by assuming $\alpha_{I}(\omega)$ is constant, and by using the Clausius-Mossotti relation, Eq.~\ref{eq:extClauMos}, as:
\begin{equation}\label{eq:thermaldragclasmos}
    \xi= \frac{32\pi^5 a^6 \hbar \beta}{135 m} \left( \frac{k_B T}{\hbar c} \right)^8 \frac{(\chi_R^2 +3\chi_R+\chi_I^2)^2+ 9\chi_I^2}{[(3+\chi_R)^2+\chi_I^2]^2} + 41.47 \frac{a^3 \hbar}{m} \left( \frac{k_B T}{\hbar c} \right)^5 \frac{3\chi_I}{(3+\chi_R)^2+\chi_I^2}.
\end{equation}
Thus, we obtained the thermal drag for magnetic dipole moments in a magnetic field as the contribution from the scattering and absorption parts. Note that $\alpha_I$ in Eq. (\ref{eq:thermaldrag}) represents the total extinction, which includes scattering and absorption contributions in Eq. (\ref{eq:withabs}). Consequently, extinction does not necessarily vanish when $\alpha(\omega)$ is purely real, such as a pure nanodiamond, due to scattering term proportional to $|\alpha(\omega)|^2$ (see Appendix \ref{sec:Extinction} and Ref. \cite{vanDeHulst1957} Sec. 6.12). This is made explicit in Eq. (\ref{eq:thermaldragclasmos}), where the first and second terms correspond to scattering and absorption, respectively. However, if the imaginary component of ${\rm Im}{\chi}=\chi_I=0$, the second term drops out completely, leaving us with the
contributions arising from ${\chi_R}$.

On the other hand, the thermal drag of a dielectric nanosphere that has the same mass within an electric field without absorption has a form \cite{Sinha:2022snc}
\begin{equation}
     \xi_E = \frac{512\pi^7 a^6 \hbar}{135 m} \left( \frac{k_B T}{\hbar c} \right)^8\left| \frac{\varepsilon-1}{\varepsilon+2} \right|^2.
\end{equation}
Therefore, we could write the symmetry between the diamagnetic nanosphere with $\beta=1/2$ and the dielectric nanosphere within their respective fields without the absorption term as 
\begin{eqnarray}\label{Ratio-drag}
     \frac{\xi_B}{\xi_E} &=& \frac{1}{32\pi^2} \left( \frac{\chi_R}{3+ \chi_R} \right)^2 \left| \frac{\varepsilon+2}{\varepsilon-1}\right|^2.
\end{eqnarray}
We will discus the physical consequences in the discussion section.

\section{Discussion}
\label{sec:iscussion}

This paper provides a complete decoherence of a nanoparticle in presence of the electromagnetic fluctuations in a finite temperature bath, where the magnetic dipole of the nanoparticle is interacting with the bath. Here, we take a point-dipole approximation, because the magnetic field gradient is not changing over the size of the particle. Our calculations are modelled for a spherical nanoparticle. In this paper, we do not take the dielectric properties of the nanodiamond, which is in fact dominant, but has already been computed in \cite{Sinha:2022snc}.
We were keen to compute the diamagnetic properties because of the levitating neutral nanodiamond which will be used for the QGEM experiment in creating the macroscopic quantum superposition for a matter-wave interferometer, which has applications for extreme acceleration sensors. In the following we summarise and discuss the physical interpretation of three main results discussed in this paper.

\noindent
{\it Momentum Diffusion Constant}:
In this paper, we improved our earlier computation of the decoherence by including the time-dependent electric field fluctuations, see~\cite{Zhang:2025fxs}. It turns out that this term dominates over the pure magnetic gradient term. Moreover, we also added the imaginary component of the magnetic polarizability, which we did not add before in \cite{Zhang:2025fxs}. Although for a pure nanoparticle, such as nanodiamond, the imaginary part of the magnetic polarizability is nearly vanishing, but it can be non-zero if dissipative from quasiparticles, and other loss channels are considered, which leads to heat loss. In a real world experiment, the imaginary part of the magnetic nanodiamond has not been measured in a levitating setup. Hence, we are unable to put any limit on decoherence due to solely the magnetic properties of the nanodiamond. However, it is paramount to know how much will be the imaginary component of the magnetic polarizability in a real life experiment, and hence we can take it into account while creating the spatial superposition in a lab. If we just considered the real part of the magnetic polarizability and only the one influenced by the magnetic field in Eq.~(\ref{eq:magp}), then we can compare our results with that of the dielectric properties of a nanodiamond by taking Eq.~(\ref{eq:dielecmdc}), and setting $\chi_I$ to be zero, the ratio is given by:
\begin{equation}\label{eq:mdcratio}
\frac{\gamma_{B}}{\gamma_{E}}=\frac{1}{16 \pi^2}\left(\frac{\chi_R}{3+\chi_R}\right)^2\left|\frac{\epsilon+2}{\epsilon-1}\right|^2
\end{equation}
As noticed earlier~\cite{Zhang:2025fxs}, 
(however, the expression in \cite{Zhang:2025fxs} was not correct, because in \cite{Zhang:2025fxs} the authors assumed large occupation number $n(\omega)(1+n(\omega))$ (see Eq.~\ref{eq:psq}), however, \cite{Sinha:2022snc} worked in  the opposite regime, so the comparison was not correct in \cite{Zhang:2025fxs}. Here, we calculate the ratio of decoherence rate due to the matter's diamagnetic and dielectric properties.
Note that the ratio is very suppressed if we ignore the imaginary component of the magnetic polarizability. For a typical solid material, $\epsilon > 1$, and for diamagnetic material, $\chi_R\ll 1$. For a pure superconductor $\chi_R =-1$;
in this case, the above ratio is $10^{-3}$
, and for a pure nanodiamond $\chi_R\sim -2.2\times 10^{-5}$ and 
$\epsilon\sim 5.7$, the ratio is $ \gamma_B/\gamma_E=9.14 \times 10^{-13}$. We do not know the ratio if we were to take the imaginary part of $\chi_v$, as it depends on the nature and concentration of the impurities present in the nanodiamond.

If we wish to create a spatial superposition size, $\Delta x \sim {\cal O}(11){\rm \mu m}$, for
masses, $m\sim {\cal O}(10^{-14})$ kg, see \cite{SchutMazumdar2025}, relevant for the QGEM experiment, and at an operating ambient temperature 
of $T \approx 5$ K, the decoherence rate given by Eq. (\ref{eq:mdcdecoherence}) is $\gamma \approx 3.16 \times 10^{-16}$ Hz. This result is valid in the long-wavelength limit since the ambient wavelength is in millimeter scale and valid within the point-dipole approximation since $|B|/|\nabla B| = \lambda/2 \pi$ is larger than the radius $a \sim {\cal O}(1){\rm \mu m}$. Furthermore, the magnetic susceptibility of nanodiamond is extremely small and absorption is negligible \cite{richards_time-resolved_2025,yelisseyev_magnetic_2009,poklonski_magnetic_2023}. Therefore, within the temperature and frequency ranges considered in this work, we approximate the magnetic susceptibility $\chi_v$ and magnetic polarizability to be frequency-independent. 
As we can see in Eq.~(\ref{eq:mdcratio}), the decoherence rate from the magnetic contribution is very suppressed and will be negligible.

Finite size effect might changes our result through multipole expansion term when the Rayleigh condition $k<<a$ is not satisfied \cite{stratton2007electromagnetic}. However, throughout this work we assume the Rayleigh regime, where the magnetic dipole term is the leading contribution.

\noindent
{\it Decoherence factor due to magnetic dipole}:
We also computed the decoherence due to the diamagnetic nanoparticle by including the imaginary part of their polarizabilities in Sec. \ref{sec:decoherence}.  If we ignore the imaginary part of the magnetic polarizability, then the decoherence rate of the dielectric sphere can be compared to that of a diamagnetic sphere by ignoring the imaginary part of the magnetic polarizability, see Eq.~(\ref{ratio-2diamag}). For a pure nanodiamond with $\chi_R\sim -2.2\times 10^{-5}$ and 
$\epsilon\sim 5.7$, the ratio is $F_B / F_E \sim 9.14 \times 10^{-13}$. This demonstrates that the decoherence rate due to the ambient electric field is much higher than the magnetic field.

\noindent
{\it Drag Force}:
Lastly, we considered the drag force on the diamagnetic nanoparticle.  Also, we compared the drag force due to the dielectric nanoparticle (see Eq.~(\ref{Ratio-drag})), again the ratio $\xi_B/\xi_E\sim 4.57 \times 10^{-13}$ for a pure nanodiamond case, suggesting that the drag force and thermal drag due to the diamagnetic interactions of a nanoparticle is much weaker than that of the dielectric interaction.
Again, if we wish to create a superpostion of a nanodiamond with a superposition size of $\Delta x \sim {\cal O}(10){\rm \mu m}$ with the decoherence rate of $\gamma\approx 3.16 \times 10^{-16}$ Hz at $T\approx 5$ K, the thermal drag given by Eq. (\ref{eq:thermaldragclasmos}) yields $\xi \approx 1.05 \times 10^{-38} ~\mathrm{Hz}$. Therefore, the thermal drag is significantly smaller and can be considered negligible compared to the decoherence rate due to the ambient magnetic field at $T\approx 5$ K.

\section*{Acknowledgments}

We would like to thank Martine Schut for insightful discussions. A.A. is supported by Lembaga Pengelola Dana Pendidikan (LPDP) of The Ministry of Finance of Indonesia. A.M.’s research is funded by the Gordon and Betty Moore Foundation through Grant
GBMF12328, DOI 10.37807/GBMF12328. This material is based on work supported by the Alfred P. Sloan Foundation under Grant No. G-2023-21130.

\appendix

\section{Statistical Independent between \texorpdfstring{$B_j$ and $\partial_i B_j$}{Bj and d_i Bj}}

Similar to \cite{Sinha:2022snc}, let us consider two classical Gaussian random processes, $X$ and $Y$, that $\langle X \rangle = \langle Y \rangle = 0$ and $\langle XY \rangle = 0$. It implies that $X$ and $Y$ are statistically independent. Then, the joint probability could be written as
\begin{equation}
    P_{XY}(X,Y) = P_X(X) P_Y(Y).
\end{equation}
Therefore, in our case, the $B_j$ and $\partial_i B_j$ are Gaussian processes, and they meet the requirement above that $\langle B_j \rangle = \langle \partial_i B_j \rangle = \langle B_j \partial_i B_j \rangle = 0 $. Therefore, we could say they are statistically independent.

\section{Derivation of \texorpdfstring{$\Delta p_E$}{pE}} \label{sec:pE}
From Eqs. [\ref{eq:pgeneral}], we could write that
\begin{equation}
    \Delta p_E = \frac{1}{c^2} \epsilon_{ijk} \int_0^{\Delta t} dt m_j \partial_t E_k.
\end{equation}
Substituting eqs. [\ref{eq:B}] and [\ref{eq:mj}], the eqs. become
\begin{eqnarray}
    \Delta p_E &=& \frac{i}{c^2} \epsilon_{ijk} \sum_{\textbf{k}_1 \lambda_1} \sum_{\textbf{k}_2 \lambda_2}  \left( \frac{\hbar \omega_1}{2 \varepsilon_0 V} \right)^{1/2} \left( \frac{\hbar}{2 \varepsilon_0 \omega_2 V} \right)^{1/2} \omega_{1} \int_0^{\Delta t} dt \left[ \alpha(\omega_2)a_{\textbf{k}_2 \lambda_2} e^{-i \omega_2 t} - \alpha^*(\omega_2) a_{\textbf{k}_2 \lambda_2}^{\dagger} e^{i \omega_2 t} \right] \nonumber \\
    & & \left[ a_{\textbf{k}_1 \lambda_1} e^{-i \omega_1 t} -  a_{\textbf{k}_1 \lambda_1}^{\dagger} e^{i \omega_1 t} \right]  (\epsilon_{ijk} \textbf{e}_{\textbf{k} \lambda,k}) (\epsilon_{jik} \textbf{k}_{i,2} \textbf{e}_{\textbf{k} \lambda,k})
\end{eqnarray}
Where the indices 1 and 2 are for $\partial_t E_k$ and $m_j$. After we ignore any non-resonance term $(\omega_1 +\omega_2)$ and do the integral to $dt$, we get
\begin{eqnarray}
    \Delta p_E &=& \frac{2i}{c^2} \sum_{\textbf{k}_1 \lambda_1} \sum_{\textbf{k}_2 \lambda_2}  \left( \frac{\hbar \omega_1}{2 \varepsilon_0 V} \right)^{1/2} \left( \frac{\hbar}{2 \varepsilon_0 \omega_2 V} \right)^{1/2} \omega_{1} \frac{\sin{[\frac{1}{2} (\omega_1 - \omega_2)\Delta  t]}}{\omega_1 - \omega_2} \times \nonumber \\
    & &  \left[ \alpha(\omega_2) a_{\textbf{k}_2 \lambda_2} a_{\textbf{k}_1 \lambda_1}^{\dagger} e^{i (\omega_1 - \omega_2) \Delta t /2} - \alpha^*(\omega_2) a_{\textbf{k}_2 \lambda_2}^{\dagger} a_{\textbf{k}_1 \lambda_1} e^{-i (\omega_1 - \omega_2) \Delta t /2} \right] ( \epsilon_{ijk} \textbf{e}_{\textbf{k}_1 \lambda_1 ,k}) \times \nonumber \\
    & & (\epsilon_{jik} \textbf{k}_{2,i}\textbf{e}_{\textbf{k}_2 \lambda_2 ,k}).
\end{eqnarray}
Where it has the exact form as Eqs. [\ref{eq:deltapE}]. 

\section{Extinction and Scattering Cross Sections of a Magnetic Dipole Moment in The Rayleigh Scattering Limit of a Lossy Particle} \label{sec:Extinction}
The energy of light beam that traverse through a medium decreases by scattering and absorption; the beam is attenuated \cite{vanDeHulst1957}. The attenuation is called as extinction. Therefore, the extinction cross section \cite{Neuringer1966,stratton2007electromagnetic} can be written as
\begin{equation}\label{eq:gencross}
    \sigma_{ext} = \sigma_{sca} + \sigma_{abs}.
\end{equation}
The scattering cross section of a magnetic dipole moment in the Rayleigh Scattering limit \cite{Neuringer1966} can be written as
\begin{equation}\label{eq:sca}
    \sigma_{sca} = \frac{6\pi}{k^2} |a_1^r|^2,
\end{equation}
and the extinction cross section is
\cite{Neuringer1966,stratton2007electromagnetic}:
\begin{equation}\label{eq:ext}
    \sigma_{ext} = \frac{6\pi}{k^2} \mathrm{Re}(a_1^r)
\end{equation}
Where $a_1^r$ is the Mie coefficient that is associated to magnetic dipoles contribution to the scattering (see \cite{stratton2007electromagnetic} page 571), which has a form given by:
\begin{equation}\label{eq:mie}
    a_1^r = \frac{i}{6\pi} \mu_0 k^3 \alpha(\omega) \beta,
\end{equation}
where $\beta$ is a dimensionless form factor, and $\beta=1/2$ for a nanosphere, and $\alpha (\omega)=\alpha_{R}(\omega)+i\alpha_I(\omega)$.
By substituting Eq. (\ref{eq:mie}) to Eqs. (\ref{eq:sca}) and (\ref{eq:ext}), we can write
\begin{equation} \label{eq:scasig}
    \sigma_{sca} = \frac{\mu_0^2}{6 \pi} \frac{\omega^4}{c^4} |\alpha(\omega)|^2 \beta^2,
\end{equation}
and
\begin{equation} \label{eq:extsig}
    \sigma_{ext} = \mu_0 \frac{\omega}{c} \alpha_I(\omega) \beta.
\end{equation}
Here, $\alpha_I(\omega)$ is the surviving term after we take the real part of $a_1^r$. If the scattering is elastic so there will be no energy absorption by the magnetic dipole, from Eqs. \ref{eq:scasig} and \ref{eq:extsig}
\begin{equation}\label{eq:opttheorem}
    \alpha_I(\omega) = \frac{\mu_0}{6 \pi} \frac{\omega^3}{c^3} |\alpha(\omega)|^2 \beta.
\end{equation}
Where it has the exact form in eq. \ref{eq:noabs}. In case there is an absorption, we must consider the absorption cross section $\sigma_{abs}$ in Eq. \ref{eq:gencross}. From \cite{Neuringer1966,stratton2007electromagnetic,vanDeHulst1957}, we could infer the absorption cross section for a magnetic dipole moment to be:
\begin{equation}\label{eq:abscross}
    \sigma_{abs} = \mu_0 \frac{\omega}{c} \alpha_I(\omega) \beta.
\end{equation}
Note that this expression has the same functional form as the extinction cross section. If the polarizability $\alpha(\omega)$ is purely real, Eq. (\ref{eq:extsig}) would vanish, which is unphysical since scattering still contributes to extinction. Therefore, $\alpha_I(\omega)$ in Eq. (\ref{eq:extsig}) should be interpreted as the extinction contribution from both scattering and absorption. On the other hand, $\alpha_I(\omega)$ in Eq. (\ref{eq:abscross}) represents only the absorption contribution, analogous to the electric dipole moment case (see~\cite{vanDeHulst1957} Sec.~6.12).

Therefore, substituting Eq. (\ref{eq:scasig}), (\ref{eq:extsig}), and (\ref{eq:abscross}) into Eq. (\ref{eq:opttheorem}), we obtain
\begin{equation}
    \alpha_I(\omega) = \frac{\mu_0}{6 \pi} \frac{\omega^3}{c^3} |\alpha(\omega)|^2 \beta + \alpha_I^{\mathrm{abs}}(\omega).
\end{equation}
Here $\alpha_I(\omega)$ comes from the extinction cross section in Eq. (\ref{eq:extsig}) as extinction coefficient and $\alpha_I^{\mathrm{abs}}(\omega)$ comes from the absorption contribution in Eq. (\ref{eq:abscross}).

\section{Momentum Diffusion Constant: No Absorption Contribution}

If we ignore the imaginary contribution to $\alpha_{I}^{\mathrm{abs}}$ in Eq. (\ref{eq:mdcwithabs}), then ${\rm Im}(\chi_v)=0$, we could obtain Eqs. (\ref{eq:pE}),(\ref{eq:pM}), and (\ref{eq:pc}) as
\begin{eqnarray}
    \frac{\langle \Delta p^2_E \rangle}{\Delta t} &=& \frac{ \mu_0^2 \hbar^2}{3 \pi^3 c^{8}} \frac{a^6}{\mu_0^2} \left(\frac{\chi_R}{3 + \chi_R} \right)^2 \left( \frac{k_B T}{\hbar} \right)^9 \zeta(8) \Gamma(9), \label{eq:elecp}\\ 
    \frac{\langle \Delta p^2_M \rangle}{\Delta t} &=& \frac{\mu_0^2 \hbar^2}{9 \pi^3 c^{8}} \frac{a^6}{\mu_0^2} \left(\frac{\chi_R}{3 + \chi_R} \right)^2 \left( \frac{k_B T}{\hbar} \right)^9 \zeta(8) \Gamma(9), \label{eq:magp}\\
    \frac{\langle \Delta p^2_c \rangle}{\Delta t} &=& -\frac{ 2 \mu_0^2 \hbar^2}{9 \pi^3 c^{8}} \frac{a^6}{\mu_0^2} \left(\frac{\chi_R}{3 + \chi_R} \right)^2 \left( \frac{k_B T}{\hbar} \right)^9 \zeta(8) \Gamma(9). \label{eq:coup}
\end{eqnarray}
Where $\zeta(8)=\pi^8/9450$, and we found a similar result for the magnetic part as in our earlier paper \cite{Zhang:2025fxs}\footnote{In our previous paper, there is a typo on momentum diffusion constant for magnetic part which it should be $\Gamma(9)$ instead of $\Gamma(8)$.}. Therefore, the total momentum diffusion constant could be written as follows
\begin{equation}
    \frac{\langle \Delta p^2 \rangle}{\Delta t} = \frac{2 \hbar^2 a^6 c}{9 \pi^3} \left(\frac{\chi_R}{3 + \chi_R} \right)^2 \left( \frac{k_B T}{\hbar c} \right)^9 \zeta(8) \Gamma(9).
\end{equation}
This result included $n^2(\omega)$ and $n(\omega)$ into account, while the $n^2$-term is usually neglected due to its small contribution.\footnote{
The factors of $n^2(\omega)$ and $n(\omega)$ are still important in determining the Planck spectrum. 
If we change $n^2(\omega) + n(\omega)$ with $n(\omega)$, we would get:
\begin{equation}\label{eq:decmdc}
    \frac{\langle \Delta p^2_M \rangle}{\Delta t} = \frac{ \hbar^2 a^6 c}{9 \pi^3}  \left(\frac{\chi_R}{3 + \chi_R} \right)^2 \left( \frac{k_B T}{\hbar c} \right)^9 \zeta(9) \Gamma(9). \nonumber
\end{equation}
The factor $n^2(\omega)+n(\omega)$ in the momentum diffusion constant corresponds respectively to the wave and particle contributions to the Einstein fluctuation formula for blackbody radiation. However, in a diamagnetic nanosphere, when we ignore $n^2(\omega)$, it has the same form as the collisional decoherence rate. It implies that it only has a particle contribution from the field.}

On the other hand, the momentum diffusion constant of a dielectric nanosphere in an electric field without absorption contribution can be written as \cite{Sinha:2022snc}
\begin{equation}\label{eq:dielecmdc}
    \frac{\langle \Delta p^2_e \rangle}{\Delta t} = \frac{16\hbar^2 a^6 c}{9 \pi} \left| \frac{\varepsilon -1}{\varepsilon + 2} \right|^2 \left( \frac{k_B T}{\hbar c} \right)^9 \zeta(8) \Gamma(9).
\end{equation}
Therefore, we could obtain the ratio between Eqs. (\ref{eq:magp}) and (\ref{eq:dielecmdc}) in the form of the decoherence rate $\gamma_B/\gamma_E$ in Eq. (\ref{eq:mdcratio}).

\section{Fokker-Planck equation}\label{sec:FokkerPlanck}

The drag force and momentum diffusion constant can be used to find the velocity distribution $w(v,t)$ of the magnetic dipoles with Fokker-Planck equation
\begin{equation}
    \frac{\partial w}{\partial v} = - \frac{\partial (Fw)}{\partial v} + \frac{1}{m}\frac{\partial^2 (D w)}{\partial v^2}.
\end{equation} 
Where $F$ is, in this case, the drag force in Eq. [\ref{eq:Findtot}] and $D$ is the momentum diffusion coefficient, where $D \equiv \frac{1}{2} \langle \Delta p^2 \rangle/\Delta t$.

We know from the momentum diffusion constant for the magnetic part in eq [\ref{eq:magp}] and $\xi$ has a fluctuation-dissipation relation
\begin{equation}
    \frac{\langle \Delta p^2 \rangle}{\Delta t} = 2 m \xi k_B T
\end{equation}
Hence, the Fokker-Planck equation, for our case, is
\begin{equation}
    \frac{\partial w}{\partial t} = -\xi \frac{\partial}{\partial v}(vw) + \frac{\xi k_B T}{m} \frac{\partial^2 w}{\partial v^2}.
\end{equation}
Where the solution, for $\xi t >> 1$, is
\begin{equation}
    w(v,t) = \left( \frac{m}{2\pi k_B T} \right)^{1/2} e^{-mv^2/2k_B T}.
\end{equation}
Therefore, it also implies that the particle velocities follow the Maxwell distribution, since the momentum diffusion constant and the drag force are subject to the assumption of $\langle \frac{1}{2} m v^2 \rangle = \frac{1}{2} k_B T$ \cite{Milne1926MaxwellsLaw}.

\bibliographystyle{unsrt}
\bibliography{References}

\end{document}